\documentclass[a4paper,11pt]{article}

\usepackage{jheppub}
\usepackage{multicol}
\usepackage{multirow}
\usepackage{subfigure}
\usepackage{slashed}
\newcommand{\met}{\ensuremath{\slashed{E}_T}}
\newcommand{\www}{{\sc $W^{\pm}W^{\pm}W^{\mp}$ }}

\newcommand{\fbinv} {\mbox{\ensuremath{\,\text{fb}^\text{$-$1}}}}

\newcommand{\pythia}{{\sc Pythia}}
\newcommand{\delphes}{{\sc Delphes}}

\newcommand{\mgme}{{\sc MadGraph/MadEvent}}

\newcommand{\madgraph}{{\sc MadGraph}}
\newcommand{\tauola}{{\sc TAUOLA}}
\newcommand{\madevent}{{\sc MadEvent}}
\newcommand{\feynrules}{{\sc FeynRules}}
\newcommand{\TMVA}{{\sc TMVA}}
\newcommand{\ssl}{$l^{\pm}\nu l^{\pm}\nu jj$}

\newcommand{\njet}{\ensuremath{N_{\mathrm{jet}}}}
\newcommand{\njetTight}{\ensuremath{N_{\mathrm{jet}}^{\mathrm{tight}}}}

\newcommand{\nlep}{\ensuremath{N_{\mathrm{lep}}}}
\newcommand{\nleploose}{\ensuremath{N_{\mathrm{lep}}^{\mathrm{loose}}}}

\title{Probing Triple-$W$ Production and Anomalous $WWWW$ Coupling at the CERN LHC and future ${\cal{O} } (100)$\,TeV proton-proton collider}

\author{Yiwen Wen$^a$,}
\author{Huilin Qu$^a$,}
\author{Daneng Yang$^a$,}
\author{Qi-shu Yan$^b$,}
\author{Qiang Li$^a$,}
\author{Yajun Mao$^a$,}

\affiliation{$^a$Department of Physics and State Key Laboratory of Nuclear Physics and Technology, \\
Peking University, Beijing, 100871, China}
\affiliation{$^b$College of Physics Sciences, University of Chinese Academy of Sciences, Beijing 100049, China and
Center for High Energy Physics, Peking University, Beijing 100871, China}

\emailAdd{wenyw@pku.edu.cn,quhl@pku.edu.cn}

\abstract{Triple gauge boson production at the LHC can be used to test the robustness of the Standard Model and provide useful information for VBF di-boson scattering measurement. Especially, any derivations from SM prediction will indicate possible new physics. In this paper we present a detailed Monte Carlo study on measuring \www production in pure leptonic and semileptonic decays, and probing anomalous quartic gauge $WWWW$ couplings at the CERN LHC and future hadron collider, with parton shower and detector simulation effects taken into account. Apart from cut-based method, multivariate boosted decision tree method has been exploited for possible improvement. For the leptonic decay channel, our results show that at the $\sqrt{s}=8(14)[100]$ TeV pp collider with integrated luminosity of 20(100)[3000] \fbinv , one can reach a significance of 0.4(1.2)[10]$\sigma$ to observe the SM \www production. For the semileptonic decay channel, one can have 0.5(2)[14]$\sigma$ to observe the SM \www production. We also give constraints on relevant Dim-8 anomalous $WWWW$ coupling parameters.}

\date{\Date}

\keywords{Triple Gauge Boson Production, Anomalous Quartic Gauge Boson Couplings, MC Simulation, LHC}

\begin{document}
\maketitle
\flushbottom

\section{Introduction}
\label{intr}
Since the beginning of the LHC era, no significant deviation from the Standard Model (SM) of particle physics has been observed. Instead, the SM has achieved great success, especially after the recent discovery of a 125-126 GeV Higgs boson in both CMS and ATLAS experiments at the LHC~\cite{1-FGianotti,2-JIncandela,3-plb:2012gu,4-plb:2012gk}. Nevertheless, we still look forward to Beyond Standard Model physics to explain some mysterious facts, such as the existence of dark matter and the electroweak-Plank scales hierarchy problem. Hence, further test on SM and searching for new physics beyond the SM become urgent quests for both theorists and experimentalists. On the other hand, the upgrade of LHC to higher collision energy and luminosity, and the promising plan for future ${\cal{O} } (100)$\,TeV proton-proton collider, make it possible to measure various `rare' SM processes, including, e.g. multi-boson productions.

To study anomalous bosonic couplings is one possible way to explore new physics. In the framework of SM, the gauge boson self-interaction is fully determined by the $SU(2)_{L} \otimes U(1)_{Y}$ gauge symmetry. Any presences of anomalous couplings may result in observable deviation from SM. To study vector boson interactions, therefore, can either further confirm the SM and the spontaneously symmetry breaking mechanism, or shed a light on new physics.

Extra contributions other than the SM predictions can be induced by possible new physics, which can be expressed in a model independent way by introducing high-dimensional operators which lead to anomalous triple or quartic gauge couplings(aTGCs or aQGCs). The explorations of aTGCs have already been done at the LEP~\cite{5-Achard:2002vd,6-Abreu:2001rpa}, Tevatron~\cite{Gounder:1999wq,Abbott:1999aj}, and later at the LHC~\cite{Aad:2011xj,Chatrchyan:2012bd} through the dibosons production. Compared with TGCs measurement, triple gauge boson production~\cite{Campanario:2008yg, Bozzi:2009ig,Bozzi:2011wwa,Bozzi:2010sj}, though suffered from lower cross sections and complicated final state topology, is essential for testing QGCs. As discussed in Ref.~\cite{Belanger:1992qh,Bosonic:2004PRD}, it is possible that the QGCs deviate from SM prediction while the TGCs do not. For instance, the exchange of extra heavy boson between vector boson can generate tree-level contributions to four gauge boson couplings while the effect on the triple gauge vertex appears only at 1-loop and is accordingly suppressed~\cite{Belanger:1992qh,Bosonic:2004PRD}.

As to aQGCs, previous Monte-Carlo(MC) and experimental studies have been carried out at e$\gamma$ and $\gamma\gamma$ colliders~\cite{eAQGC:1993,AAQGC:1995}, linear colliders~\cite{Belanger:1992qh,zzzwwz:1996,Belanger:1999,wwaLEP:1999,zaaLEP:1999,wwaDELPHI:2003,wwaLEP:2004}, and hadron colliders\cite{Bosonic:2004PRD,AQGC:2001,Royon:2010tw,AALHC:2003,QGCZ:1996,QGC:1995,WVgammaCMS:2014,
gammagammatowwCMS:2013,samesignwwATLAS:2014}. Many experiments gave direct constraints including LEP, by studying the $WW\gamma$~\cite{wwaLEP:1999,wwaDELPHI:2003}, $Z\gamma\gamma$~\cite{zaaLEP:1999} and $\gamma\gamma\nu\bar{\nu}(q\bar{q})$~\cite{wwaLEP:2004} channels, e.g., constraints on $WW\gamma\gamma$ aQGC parameters are given. Recently, CMS presents new results on $WW\gamma\gamma$ and $WWZ\gamma$ aQGCs by studying the semi-leptonic $WV\gamma$ production~\cite{WVgammaCMS:2014} and $\gamma\gamma\rightarrow WW$ channel~\cite{gammagammatowwCMS:2013}. ATLAS has studied $WWWW$ aQGC via the same sign $WW$ channel~\cite{samesignwwATLAS:2014}.

In the next few years, the LHC at CERN will be upgraded with higher center-of-mass energy and luminosity and it is expected that it will set more strict constraints on aQGCs. The MC studies on $W^+W^-\gamma$~\cite{Daneng:2013} production and $W^{\pm}Z\gamma$ production~\cite{KeYe:2013} have confirmed the potential of LHC on probing $WW\gamma\gamma$ and $WWZ\gamma$ aQGCs. As to $WWWW$ vertices, Eboli {\it et al.}~\cite{VBFWWWW:2006} studied on the vector boson fusion(VBF) WW channel and set the aQGC parameter $f_{S0,S1}$ constraints at the order of $10^{-11}$GeV$^{-4}$ at 99\% CL with integrated luminosity of 100 \fbinv . Preliminary result from Snowmass~\citep{snowmass:2013} via the \www production pure leptonic channel set the aQGC parameter $f_{T0}/\Lambda^4$  at the order of $10^{-12}$ GeV$^{-4}$ at 5$\sigma$ with 300\fbinv 14 TeV LHC . 

This paper will present detailed study on triple gauge boson production via exploring the potential of measuring $W^{\pm}W^{\pm}W^{\mp}$ final states with full leptonic decay and semileptonic decay at the $\sqrt{s}=8$ and 14 TeV CERN LHC and future proton-proton collider, and probing the $WWWW$ anomalous coupling. Our work extends Eboli {\it et al.} and Snowmass' study as an independent test on triple electroweak gauge boson physics. We begin by introducing the aQGC related effective theory and specifying the effective Lagrangian in Sec.~\ref{effwwww}, and then present our MC simulation on SM $WWW$ production with pure leptonic decay channel in Sec.~\ref{fulllep} and semileptonic channel in Sec.~\ref{semilep}. In Sec.~\ref{anowww}, we demonstrate the result on the $WWWW$ aQGC study. The $WWW$ production and aQGC at 100 TeV hadron collider analysis will be given in Sec.~\ref{100TeV}.  Unitarity safety on the aQGC limits is discussed in Sec.~\ref{UV}.  Finally, we draw our conclusion in Sec.~\ref{conclu}.

\section{Effective Interactions for aQGCs}
\label{effwwww}
An effective Lagrangian can be constructed in a model independent way for the anomalous quartic couplings,  assuming that new physics beyond the SM still keeps $SU(2)_{L} \otimes U(1)_{Y}$ gauge invariance. The Lagrangian can be expressed in non-linear or linear representation~\cite{Belanger:1992qh,Belanger:1999}. Since a higgs boson has been discovered in LHC, it is more preferable to work in the linear context.

The lowest order genuine aQGC operators in linear representation are dimension-8(dim-8). There are three classes of such operators: operators containing only covariant derivative of the field $D_\mu\Phi$, operators containing $D_\mu\Phi$ and field strength, and operators containing only the field strength~\cite{VBFWWWW:2006}. In our research, we choose to study the below three dim-8 operators:
\begin{eqnarray}\label{oper1}
{\cal L}_{S0} = \frac{f_{S0}}{\Lambda^4}[(D_\mu\Phi)^\dag D_\nu\Phi]\times[(D^\mu\Phi)^\dag D^\nu\Phi],
\end{eqnarray}
\begin{eqnarray}\label{oper2}
{\cal L}_{S1} = \frac{f_{S1}}{\Lambda^4}[(D_\mu\Phi)^\dag D^\mu\Phi]\times[(D_\nu\Phi)^\dag D^\nu\Phi],
\end{eqnarray}
\begin{eqnarray}\label{oper3}
{\cal L}_{T0} = \frac{f_{T0}}{\Lambda^4}\mathrm{Tr}[\hat{W}_{\mu\nu}\hat{W}^{\mu\nu}]\times\mathrm{Tr}[\hat{W}_{\alpha\beta}\hat{W}^{\alpha\beta}],
\end{eqnarray}
where $f_{S0,S1,T0}$ represents the dimensionless numerical coefficients. $\Lambda$ is a mass-dimension parameter associated with the energy scale of the new degrees of freedom.
\begin{figure}{
\centering
\subfigure[Unitarity bounds up to 14 TeV]{
    \label{operator:a}
    \includegraphics[width=0.45\textwidth]{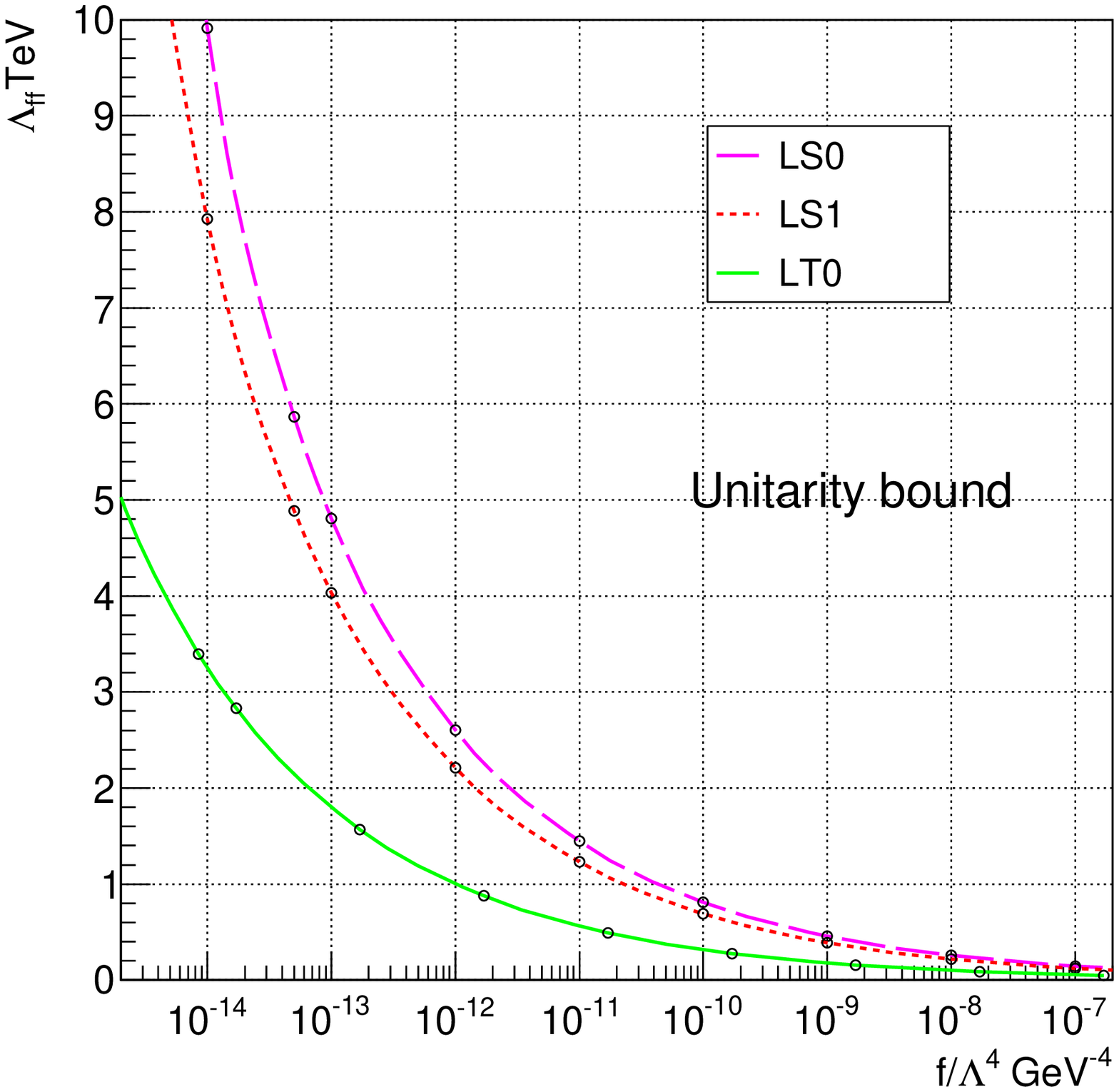}}
\subfigure[Unitarity bounds up to 100 TeV]{
    \label{operator:b}
    \includegraphics[width=0.45\textwidth]{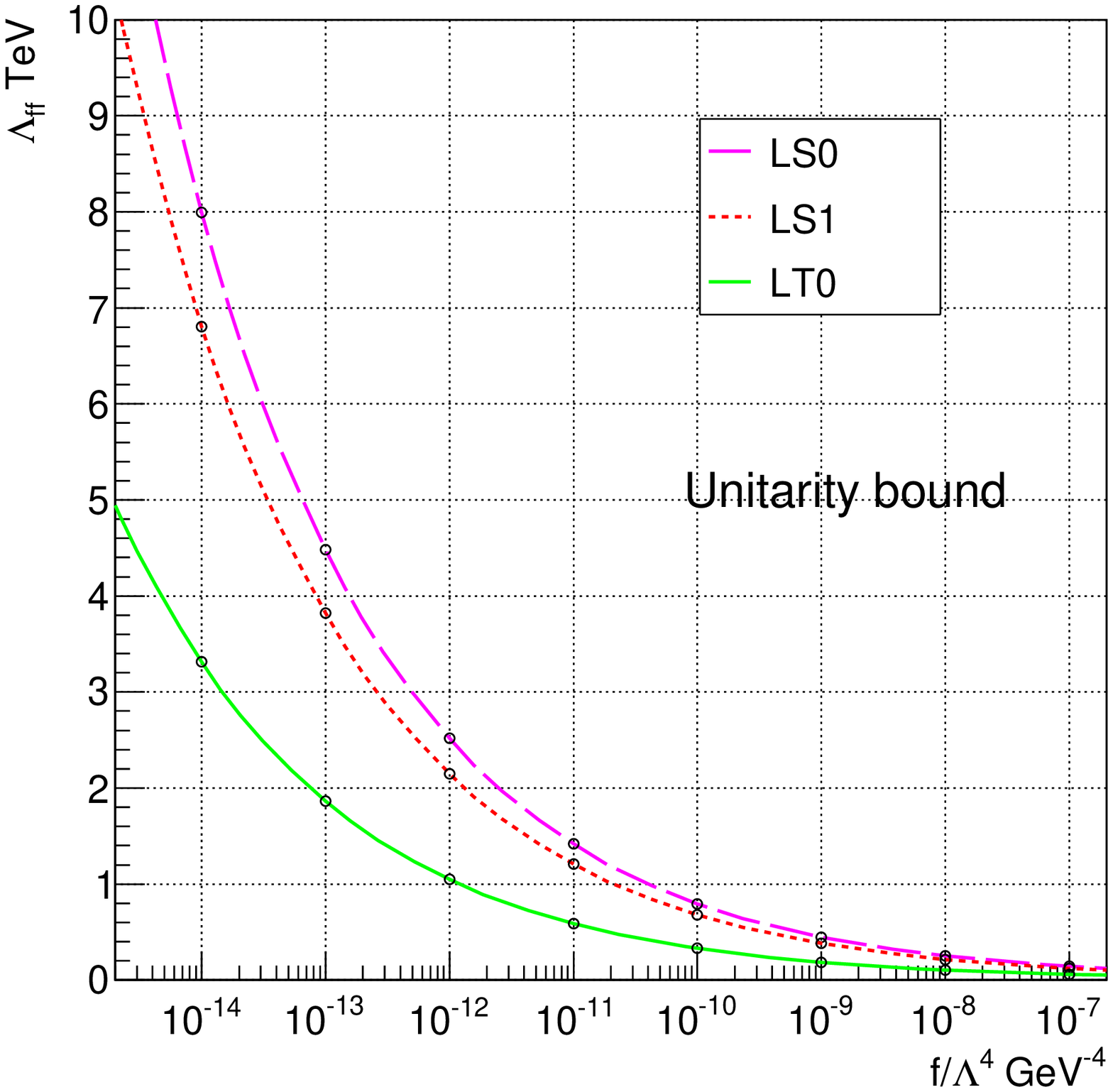}}
\caption{\label{operators} Unitarity bounds on form factor scale $\Lambda_{ff}$ for different operators, got from $2\rightarrow 2$ di-boson scattering for energy up to 14 and 100 TeV for $n=2$. Unitarity safe region is below the curves.}}
\end{figure}

One should note that the effective Lagrangian leads to tree-level unitarity violation at corresponding high energy. Usually, one can adjust the rising cross section by introducing an appropriate form factor. However, the choice of form factor is arbitrary and can be disputable~\cite{Wudka:1996ah,Maestre:2011}. In this paper, we just present our results with some typical form factor, following the commonly used formalism~\cite{Bosonic:2004PRD}.
\begin{eqnarray}\label{formf}
f_J\rightarrow\frac{f_J}{(1+\hat{s}/\Lambda^2_{ff})^n}
\end{eqnarray}
where $f_J$ could be the $f_{S0}$, $f_{S1}$ and $f_{T0}$, $\hat{s}$ is the partonic center-of-mass energy, $\Lambda_{ff}$ represents the form factor cutoff scale. 

Figs.~\ref{operators} shows the energy scale at which tree-level unitarity would be violated without a form factor ($\Lambda_{ff}\rightarrow 0$) and the form factor scale ($\Lambda_{ff}$) that ensures tree-level unitarity up to the given energy. The region below the red line is unitarity safe. These bounds are estimated by using the form factor tool available with VBFNLO\cite{VBFNLO}. The form factor is determined by calculating on-shell vector bosons scattering and computing the zeroth partial wave of the amplitude. As unitarity criterion the absolute value of the real part of the zeroth partial wave has to be below 0.5. 

\section{Standard model \www production in pure leptonic decay channel}
\label{fulllep}
The MC simulations are carried out within \mgme\ v5\cite{mg5}. The effective Lagrangian of $WWWW$ aQGCs are incorporated in \madgraph\ based on the \feynrules-UFO-ALOHA\cite{Christensen:2008py,Degrande:2011ua,deAquino:2011ub} framework. The signal and background processes are first generated at parton level by \madgraph ~\cite{mg5}  and \madevent ~\cite{Maltoni:2002qb}, and then passed through the interface to \pythia\ 6~\cite{pythia} for parton shower and hadronization. The detector simulations are done by using \delphes\ 3.0 package~\cite{Delphes}, where we focus on CMS detector at the 8 and 14 TeV LHC and a combined ATLAS-CMS detector~\cite{Snowmass:detector} at the future 100 TeV proton-proton collider. Finally, all events are delivered to {\sc ExRootAnalysis}~\cite{ExRootAnalysis} and analyzed with ROOT~\cite{root}. In the analysis step, we use both traditional cut-based method and Multivariate Analysis(MVA) boosted decision tree(BDT) method~\cite{BDT}. The MVA BDT method is carried out under the \TMVA\ package~\cite{TMVA2007} included in ROOT.

\begin{figure}{
\centering
\subfigure[With TGC]{
    \label{diagram:a}
    \includegraphics[width=0.4\textwidth]{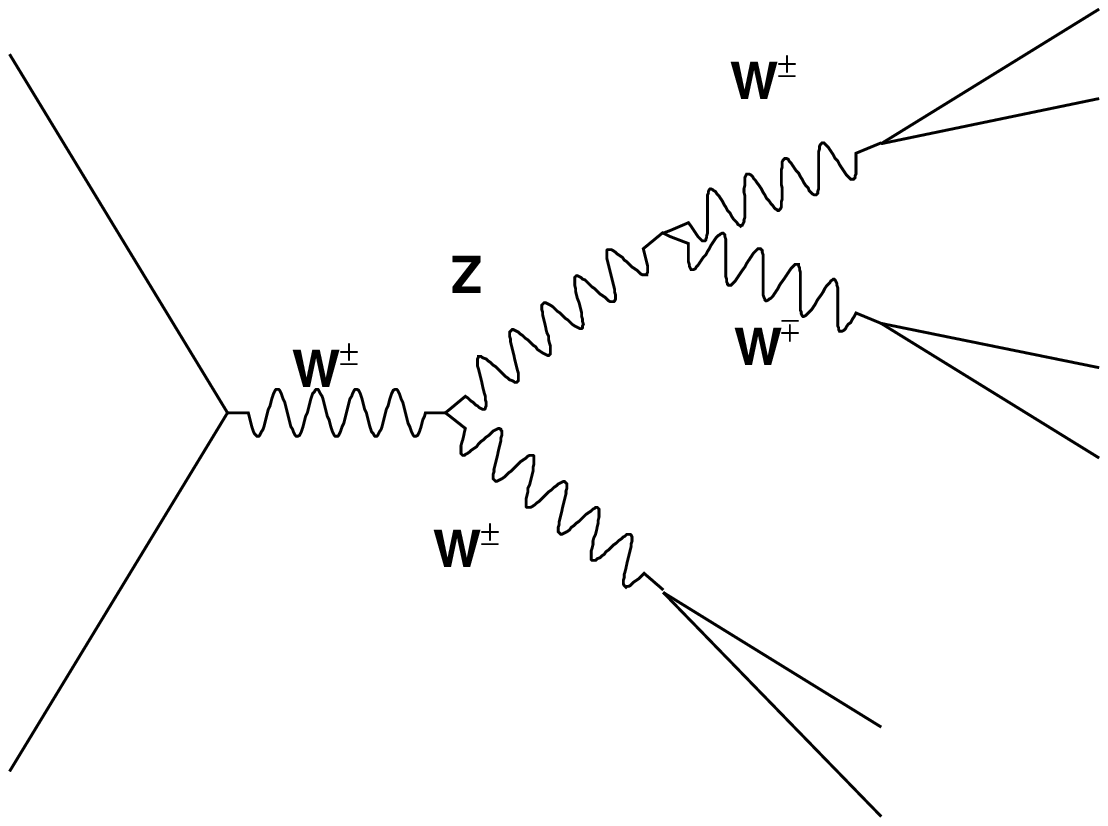}}
\subfigure[With (anomalous) QGC]{
    \label{diagram:b}
    \includegraphics[width=0.4\textwidth]{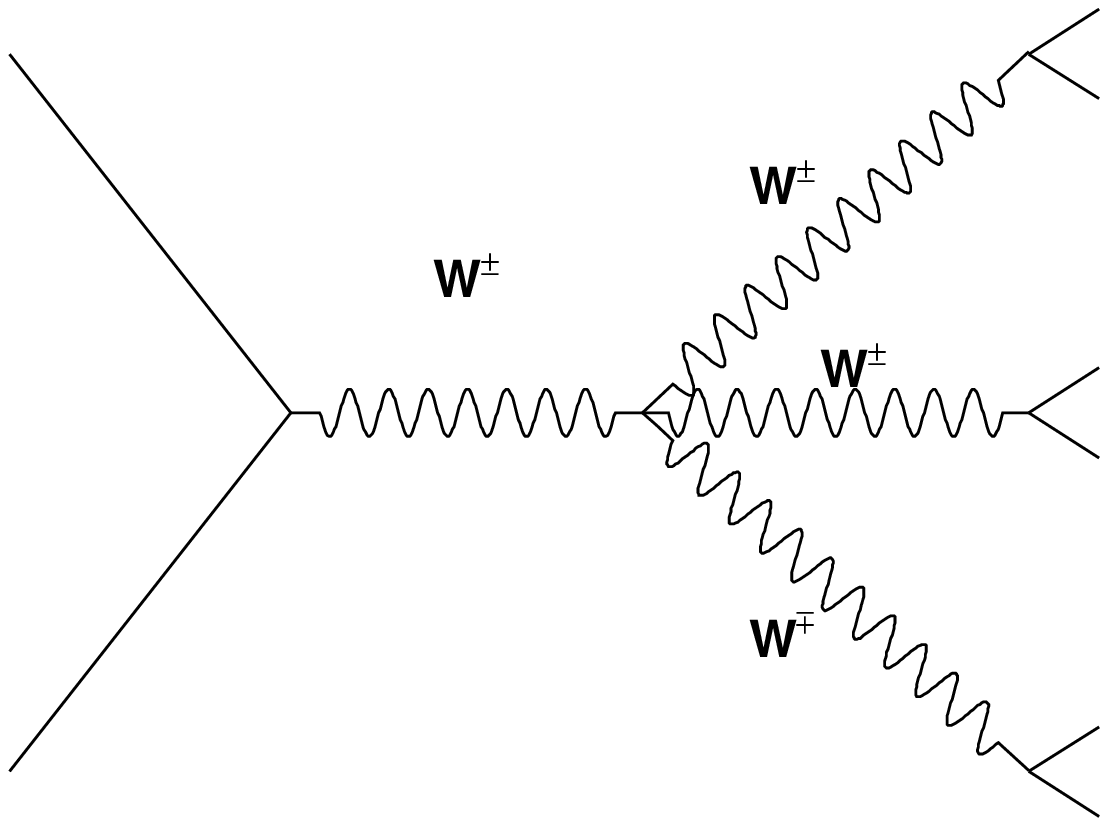}}
    \subfigure[With higgs coupling]{
    \label{diagram:c}
    \includegraphics[width=0.4\textwidth]{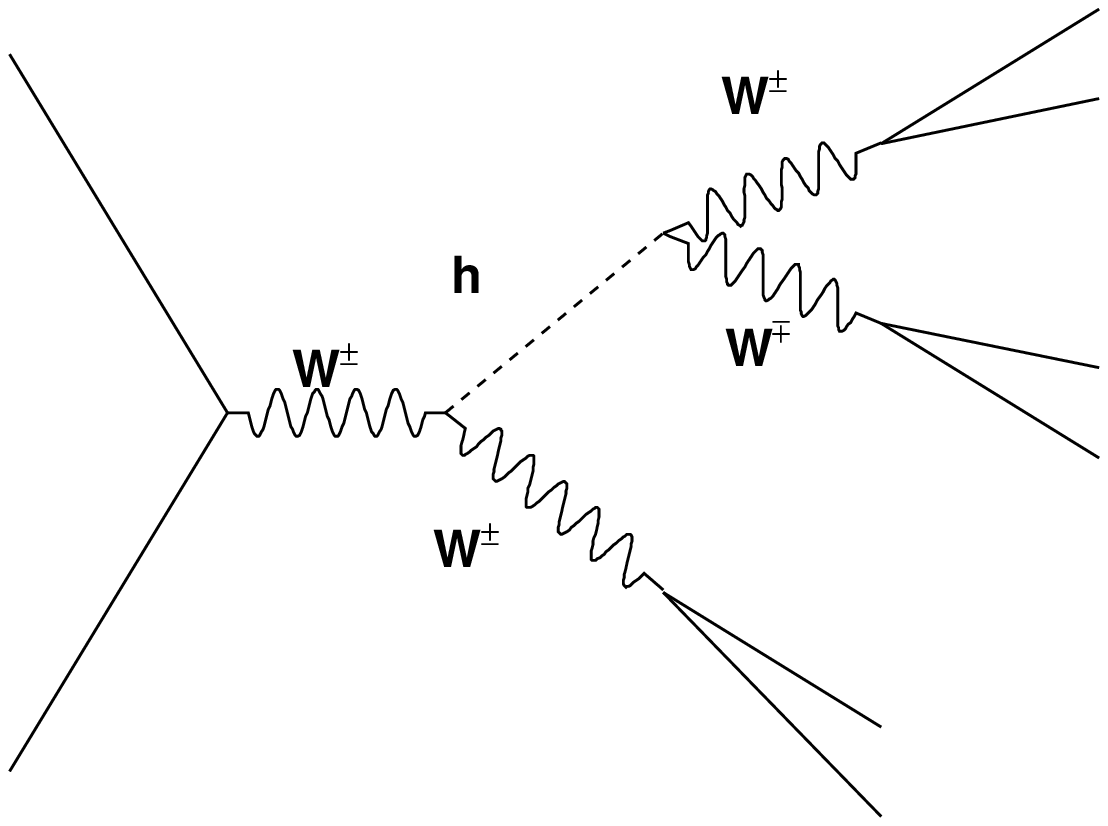}}
\subfigure[$W$ emission]{
    \label{diagram:d}
    \includegraphics[width=0.4\textwidth]{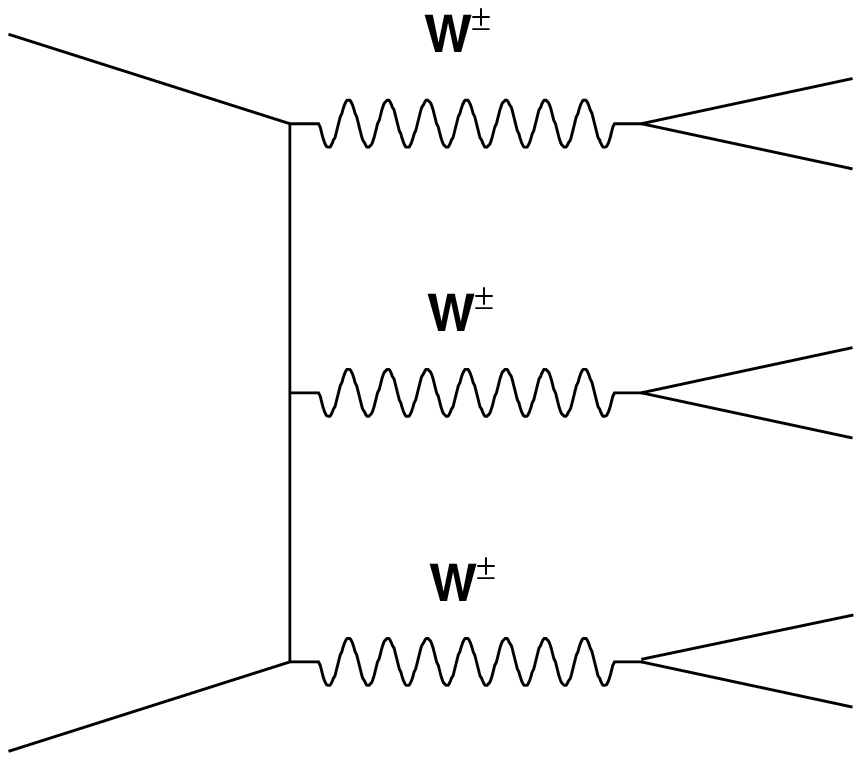}}
\caption{\label{diagrams} Example Feynman diagrams contributing to $WWW$ productions at the LHC}}
\end{figure}

The characteristic signal of this channel contains three well-defined leptons with total electric charge $\pm$1, in association with large missing transverse energy \met. Some example Feynman diagrams are plotted in Fig.~\ref{diagrams}, for $W^{\pm}W^{\pm}W^{\mp}$ production at the LHC, in the trileptonic final state $lL\bar{L}\nu$, with $l,L, \bar{L}= e,\mu$ and $\tau$. Note that $\tau$ decays into $e$, $\mu$ at the ratio of about 35\% and is handled by \tauola ~\cite{tauola}. Fig.~\ref{diagram:a} involves TGCs and Fig.~\ref{diagram:c} involves higgs coupling, both are not sensitive to aQGC.

Five main backgrounds are taken into account: $WZ$ (including virtual photon contributions), $t\bar{t}W$, $ZZ$, $t\bar{t}Z$ and $WWZ$, where $WZ$ and $t\bar{t}W$ are dominant. Notice that the 4 leptons final state can be possible backgrounds with one lepton unidentified.

In order to improve event generating efficiency, we choose the following pre-selection cuts to generate unweighted events at parton level with \mgme\ .
\begin{itemize}{
\item  (1) $P_{T\,l,j} \geq 10 $ GeV,
\item  (2) $\met \geq 20$ GeV,
\item  (3) $|\eta_j| < 5, |\eta_l| < 2.5$,
\item  (4) $R_{jj} > 0.4, R_{ll} > 0.3, R_{jl} > 0.3$,
}\end{itemize}
where $R\equiv\sqrt{{\Delta\phi}^2+{\Delta\eta}^2}$ in which $\phi$ being the azimuthal angle and $\eta$ the pesudo-rapidity of a particle. For those backgrounds containing unidentified leptons, we do not apply any of the above cuts on leptons in order not to make bias.

Meanwhile, in the hard process generation with \mgme\, we adopt the CTEQ6L1 parton distribution functions (PDFs)~\cite{Pumplin:2002vw} and set the renormalization and factorization scales as default dynamic scales.

In \delphes, we consider no pileup and  mean 20 pileup scenarios at 8 TeV LHC, no pileup, pileup 50 and 140 at 14 TeV LHC, and 50 and 140 pileup at future 100 TeV proton-proton collider.  

As mentioned before, we present both cut-based method and BDT method to evaluate the feasibility of observing $WWW$ production.
\subsection{Cut-based method}
In the cut-based analysis step, we apply the following high level cuts:
\begin{itemize}{
\item (1) In order to select signal-like events, we require 3 and only 3 leptons in one event, the sum of electric charge of 3 leptons should be 1 or -1, and leading lepton $P_{Tl}>35$ GeV, the rest two leptons' $P_{Tl} > 20$ GeV,
\item (2) $\met > 25$ GeV,
\item (3) In order to suppress top quark related backgrounds, we exclude events with b-tagged jet,
\item (4) To reject virtual photon production and leptons from hadron decay, we require the invariant mass of lepton pair $m_{ll}>12$ GeV,
\item (5) The transverse mass of 3 lepton system $m_T>300$ GeV,
\item (6) $R_{ll} > 0.5$.

}\end{itemize}
We present our analysis in two schemes.
\begin{itemize}{
\item \textit{scheme 1}(s1): We require at least one pair of opposite-sign same-flavor(OSSF) leptons and its mass $m_{OSSF}$ satisfied $|m_{OSSF}-M_Z|>15$ GeV,
\item \textit{scheme 2}(s2): We only collect events from the remaining lepton topologies: $e^-\mu^+\mu^+$, $e^+\mu^-\mu^-$, $\mu^-e^+e^+$, and $\mu^+e^-e^-$. These final states topologies only occur in triple W boson production related samples.
}
\end{itemize}
Scheme 2 is to further suppress the backgrounds with Z boson leptonic decay.

\subsection{Multivariate analysis BDT method}
In our research, we use MVA classification to study the feasibility of $W^{\pm}W^{\pm}W^{\mp}$ production. A typical MVA classification analysis consists of two independent phases: the training phase, where the MVA methods are trained, tested and evaluated, and application phase, where the methods are applied to the concrete classification problem they have been trained for.

Before going into training phase, we preselect the events with the preselection cuts:
\begin{itemize}{
\item (1) 3 and only 3 leptons in one event, the sum of electric charge of 3 leptons should be 1 or -1,
\item (2) $\met > 25$ GeV,
\item (3) Exclude events with b-tagged jet,
\item (4) the mass of arbitrary two leptons $m_{ll}>12$ GeV,
\item (5) $R_{ll}$ is larger than 0.5.
}
\end{itemize}
After preselection, we input the following discriminating variables to the TMVA package: 3 lepton's $P_{Tl}$, $\eta_l$, $R_{ll}$, transverse missing energy, transverse mass of 3 leptons $m_T$, $H_T$, $M_{lll}$, $P_{T_{lll}}$, and the $m_{ll}$ of the lepton pair with $m_{ll}$ closest to $M_Z$. These variables would be used for MVA training.

\subsection{Numerical Results}
What we are interested in is evaluating the feasibility of observing triple gauge boson $W^{\pm}W^{\pm}W^{\mp}$ Production. To optimize the results, we introduce a further requirement $P_{Tj}>P_{Tj}^{cut}$ and $n_{j}=0$ in addition to all the cuts mentioned above in cut-based analysis, where $P_{Tj}^{cut}$ is the jet reconstructing cut and $n_j$ is reconstructed jet number. The purpose of setting this special cut is to suppress more top-quark related background and keeping high signal efficiency at the same time. Compared with the signal process, $t\bar{t}W$ and $t\bar{t}Z$ tend to radiate more jets. Furthermore, the hard physics scale is higher and thus one or more jets can be harder than the jets in signal.

\begin{figure}{
\centering
\includegraphics[width=0.49\textwidth]{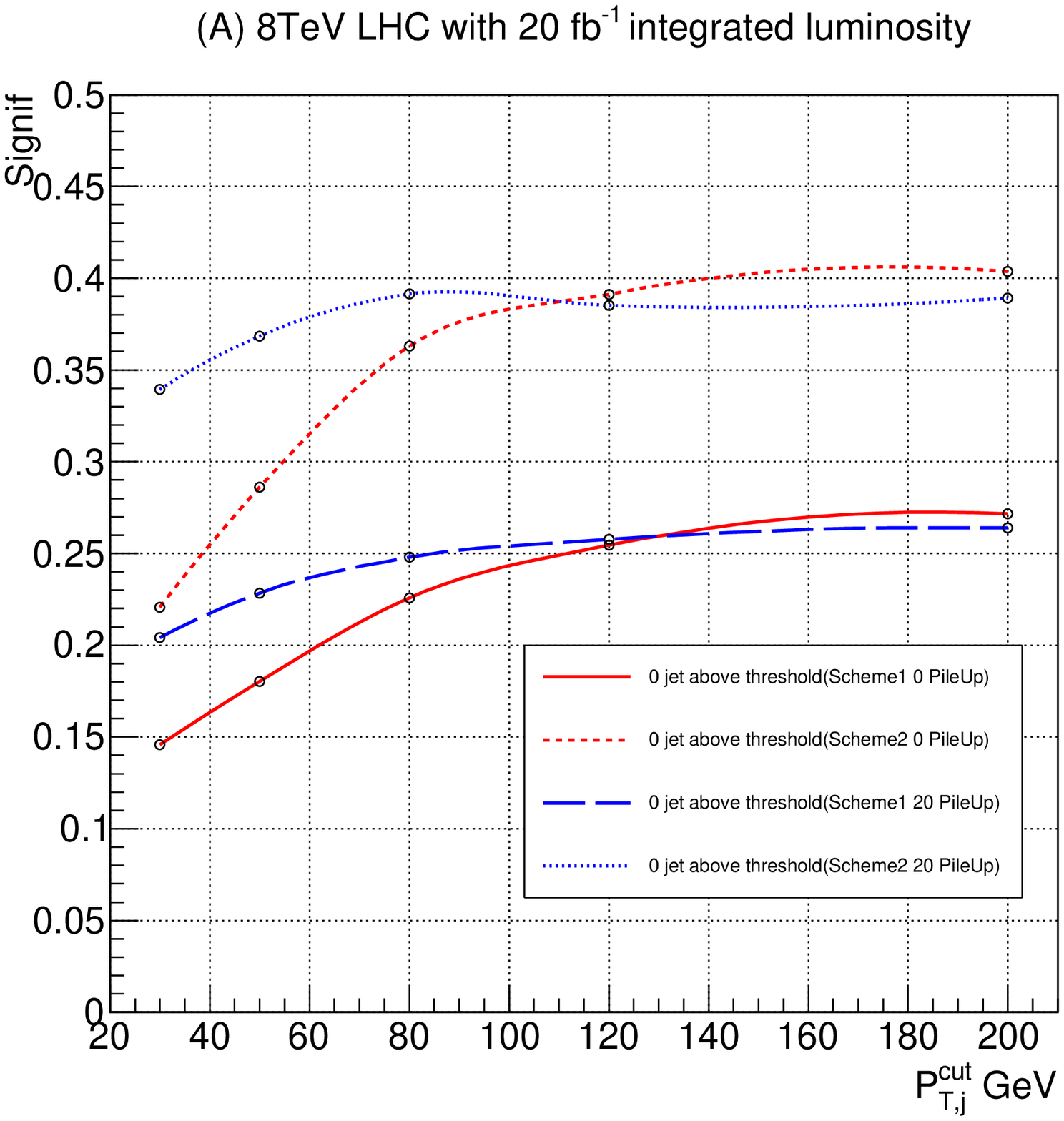}
\includegraphics[width=0.49\textwidth]{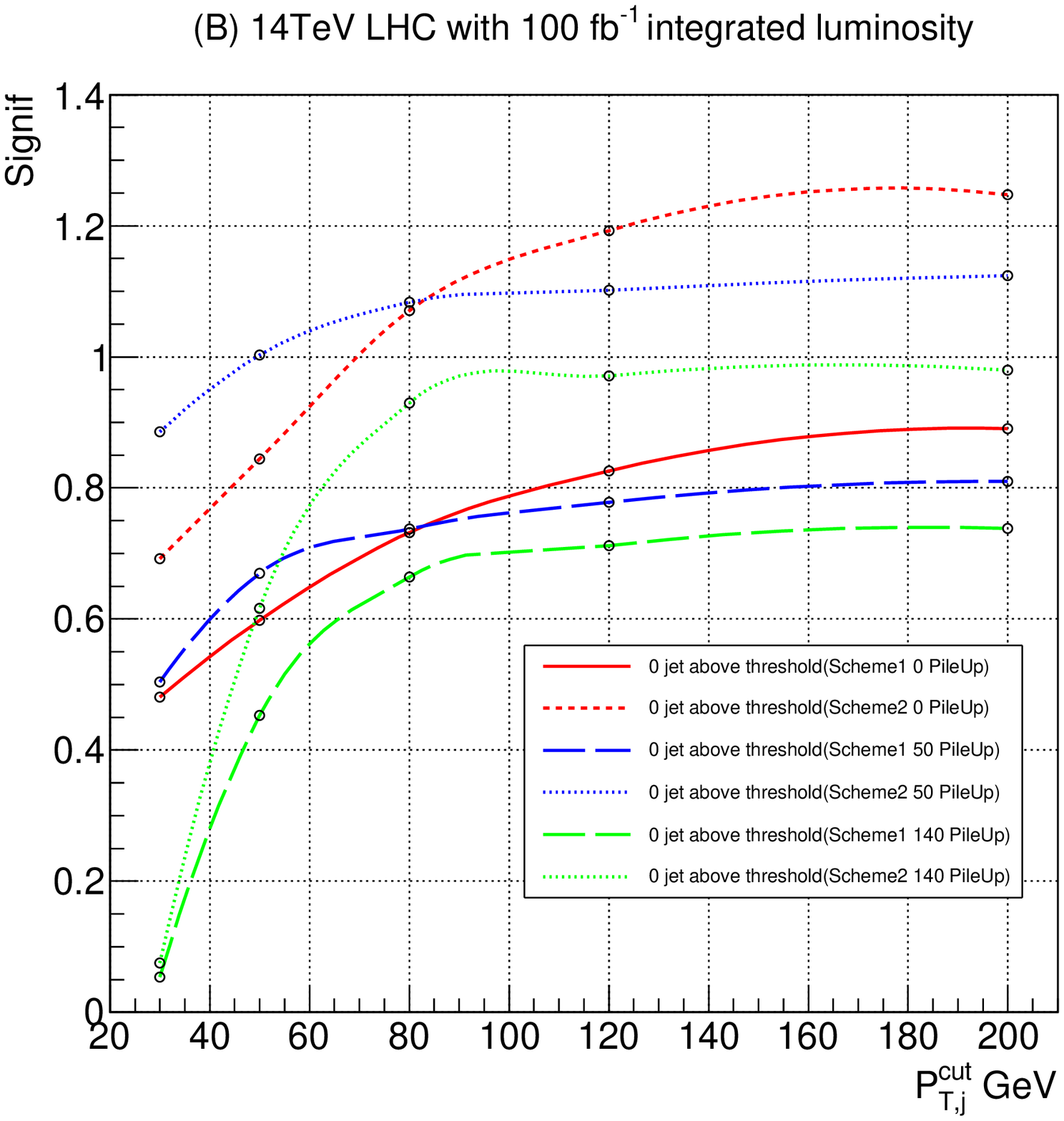}
\caption{\label{wwwcut}$WWW$ production in pure leptonic decay significances, varying jet reconstructing cut $P^{cut}_{T\,j}$.}}
\end{figure}

The significances are shown in Fig.~\ref{wwwcut}, calculated with Eq.~(\ref{stat}). It is interesting to note that when $P_{Tj}^{cut}$ is getting larger, the significance goes higher. This means the cases of no requirement of jet $P_{T}$ have the largest significances (which comes from the statistic increasing). Thus we are not going to apply any jet veto in the following.

We list the 8 TeV event numbers for the signal and backgrounds and significances in Table.~\ref{tab1} and 14 TeV in Table.~\ref{tab2}. A significance about $ 0.28\sim 0.41\sigma$ can be achieved to observe \www production in pure leptonic decay channel at 8 TeV LHC and $ 0.75\sim 1.28\sigma$ at 14 TeV LHC. We note that Scheme 2 tends to have larger significance than Scheme 1, due to further suppression on WZ background. The results of the two scheme can be combined in future experimental studies. Moreover, the BDT method can gives us some gain but not much.

\begin{eqnarray}\label{stat}
Signif = \sqrt{2 ln(Q)}, \text{} Q = (1 + N_s/N_b)^{N_{obs}} exp(-N_s).
\end{eqnarray}

\begin{center}
\begin{table*}[h!]
\begin{tabular}{c||c||c|c|c|c|c}
\hline
\multirow{4}{*}{Processes} & \multirow{4}{*}{Cross section[fb]} & \multicolumn{5}{c}{ Events } \\
\cline{3-7}
& & \multicolumn{4}{c|}{cut-based} & BDT \\
\cline{3-7}
& & \multicolumn{2}{c|}{Pileup 0} & \multicolumn{2}{c|}{Pileup 20} &Pileup 0\\
\cline{3-7}
& & s1 & s2 & s1 & s2 & s1\\
\hline
$WWW$        & 1.06     & 2.08      & 0.636   & 2.03      & 0.612 & 1.99  \\\hline
$WZ$         & 235     & 47.3      & 0.800   & 45.7      & 0.925 & 38.0  \\\hline
$t\bar{t}W$  & 3.97     & 2.81      & 0.859   & 2.96      & 0.935 & 4.84  \\\hline
$ZZ$         & 129     & 3.97      & 0.206   & 7.95      & 0.155 & 2.94  \\\hline
$t\bar{t}Z$  & 1.41     & 0.521     & 0.146   & 0.541      & 0.147 & 0.958  \\\hline
$WWZ$        & 0.358    & 0.344     & 0.0983  & 0.334     & 0.0936& 0.320  \\\hline\hline
 \multicolumn{2}{c||}{significance}& 0.279  & 0.418  & 0.266 & 0.391 & 0.288\\\hline
\end{tabular}
\caption{\label{tab1} Event numbers and significances of $WWW$ production in pure leptonic decay channel at the LHC with $\sqrt{s}=8$ TeV and integrated luminosity of $20\fbinv$.}
\end{table*}
\end{center}

\begin{center}
\begin{table*}[h!]
\begin{tabular}{c||c||c|c|c|c|c|c|c}
\hline
\multirow{4}{*}{Processes} & \multirow{4}{*}{Cross section[fb]} & \multicolumn{7}{c}{ Events } \\
\cline{3-9}
& & \multicolumn{6}{c|}{cut-based} & BDT \\
\cline{3-9}
& & \multicolumn{2}{c|}{Pileup 0} & \multicolumn{2}{c|}{Pileup 50} &\multicolumn{2}{c|}{Pileup 140} &Pileup 0\\
\cline{3-9}
& & s1& s2 & s1& s2 &s1& s2 & s1\\
\hline
$WWW$        & 2.17 & 21.0   &  6.29   & 20.0     & 5.82     & 17.9      & 5.18      & 20.2     \\\hline
$WZ$         & 412 & 421   &  6.86    & 429     & 6.72     & 398      & 6.59      & 337     \\\hline
$t\bar{t}W$  & 9.88 & 33.4   &  10.3   & 38.2     & 11.5     & 38.8      & 11.9      & 56.0     \\\hline
$ZZ$         & 273 & 40.4   &  1.09   & 98.8     & 1.64     & 107      & 2.73      & 32.7     \\\hline
$t\bar{t}Z$  & 6.35 & 10.8   &  2.78   & 12.6     & 3.46     & 13.3      & 3.60      & 18.5     \\\hline
$WWZ$        & 0.849& 3.73   &  1.04   & 3.73     & 1.01     & 3.54       & 0.949     & 3.23     \\\hline\hline
 \multicolumn{2}{c||}{significance}& 0.922   & 1.28 & 0.822 & 1.14 & 0.751 & 0.989 & 0.946 \\\hline

\end{tabular}
\caption{\label{tab2} Event numbers and significances of $WWW$ production in pure leptonic decay channel at the LHC with $\sqrt{s}=14$ TeV and integrated luminosity of $100\fbinv$.}
\end{table*}
\end{center}

\section{Standard model \www production in semileptonic decay channel}
\label{semilep}
In semileptonic decay channel analysis, we use exactly the same simulation framework as in pure leptonic decay channel in Sec~\ref{fulllep}. We only consider the most significant case \ssl\ final state here. The same sign leptons  come from the two same sign W bosons decay. The other opposite charge W boson decays hadronically into two jets. 
Here three main categories of background processes contribute to this channel: 
\begin{itemize}
\item $W^{\pm}W^{\pm}jj$ decaying to \ssl\, where the two jets don't come from an on-shell W boson. Both electroweak process and QCD process are included,
\item $WZjj$ decaying to $l^{\pm}\nu l^{\pm} l^{\mp}jj$, where one of the leptons is missing. Both electroweak process and QCD process are included,
\item $t\bar{t}W$, all decay modes are included. 
\end{itemize}

When generating events at \madgraph\ , we apply the following preselection cuts.
\begin{itemize}{
\item  (1) $P_{T,j} \geq 20 $ GeV, $P_{T,l} \geq 10 $ GeV.
\item  (2) $\met \geq 10$ GeV.
\item  (3) $|\eta_j| < 5, |\eta_l| < 2.5$.
\item  (4) $R_{jj} > 0.4, R_{ll} > 0.4, R_{jl} > 0.4$.
}\end{itemize}
Note that Backgrounds containing unidentified leptons don't have cuts related to leptons applied on.

\subsection{Cut-based method}
In the cut-based analysis step, the optimized event selection is shown in Table.~\ref{tab:ss-cut}, where \nlep\ is the number of lepton that has $P_T>20$ GeV and $|\eta|\leq 2.4$, \nleploose\ is the number of lepton that has $P_T>10$ GeV and $|\eta|\leq 2.4$, \njet\ is the number of jet that has $P_T>30$ GeV and $|\eta|\leq 5$, $m_{jj}$ is the invariant mass of the leading two jets and $m_W$ is mass of W boson. 
\begin{table}[h]
  \centering

    \begin{tabular}{ccccc}
    \hline
          & 8\,TeV  & \multicolumn{3}{c}{14\,TeV}  \\
    Pileup & 20    & 0     & 50    & 140 \\
	\hline
    \nlep  & \multicolumn{4}{c}{$=2$} \\
    \nleploose & \multicolumn{4}{c}{$=2$} \\
    lepton sign & \multicolumn{4}{c}{$(+,+)$ or $(-, -)$} \\
    \met   & \multicolumn{4}{c}{$\ge30$\,GeV} \\
    \njet  & $=2$    & $=2$    & $\ge 2, \le 3$ & $\ge 2$  \\
    
    $|m_{jj}-m_W|$ & $\le 15$\,GeV & $\le15$\,GeV & $\le20$\,GeV & $\le30$\,GeV \\\hline
    \end{tabular}%
      \caption{\ssl\ Event selections, for \njet\ and $m_{jj}$ optimization check also Figures.~\ref{distributionsemia} and ~\ref{distributionsemib}.}
  \label{tab:ss-cut}%
\end{table}%

\subsection{Multivariate analysis BDT method}
The event preselections before going into training phase of BDT are shown as below:
\begin{itemize}{
\item (1) The event contains two and only two reconstructed leptons with $P_T>20$ GeV and $|\eta|<2.4$. Events with extra lepton whose $P_T>10$ GeV and $|\eta|<2.4$ are vetoed,
\item (2) $\met > 30$ GeV,
\item (3) At least 2 jets, whose $P_{T,j} \geq 30$ GeV, $|\eta|\leq 5$.
}
\end{itemize}
The following discriminate variables will be put into TMVA packages: $P_{Tj}$, $\eta_j$, $m_j$, $N_{jet}$, leading lepton $P_T$, \met\ , the invariant mass of two leading jets $m_{jj}$, distance between the leading two jets $R_{jj}$, the azimuthal angle between two leading jets $\Delta\phi_{jj}$, lepton pair's $P_T$, $ R_{ll}$, the azimuthal angle between two leptons $\Delta \phi_{ll}$, the azimuthal angle between \met\ and the lepton pair $\Delta\phi_{ll,\met}$, the minimum of the azimuthal angles between lepton and \met\ $\Delta\phi_{l,\met}^{min}$ and the minimum distance between the leading jet and lepton $R_{jl}^{min}$.

\subsection{Numerical results}
\begin{figure}{
\centering
\subfigure[Number of jet]{
    \label{distributionsemia}
    \includegraphics[width=0.46\textwidth]{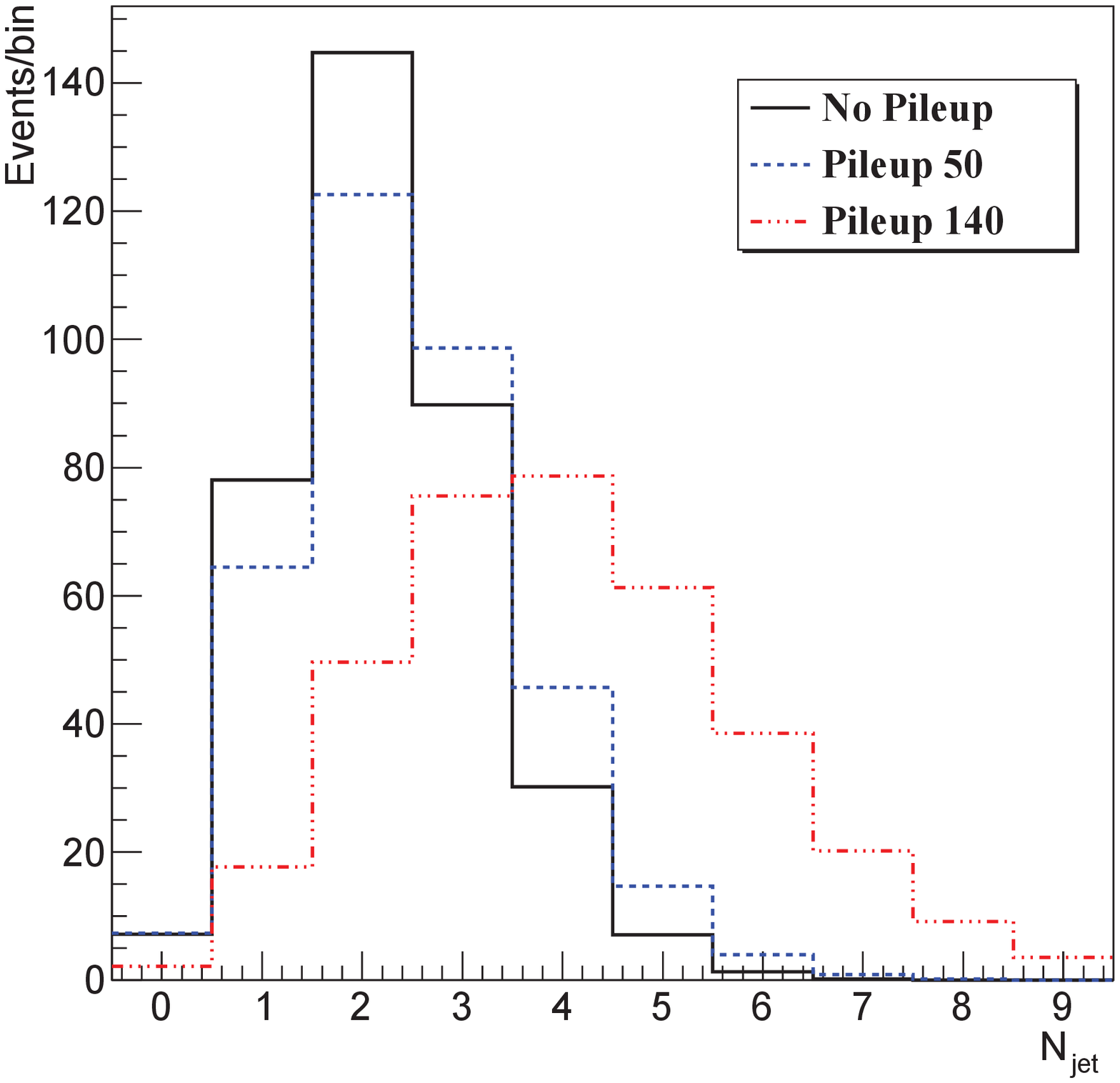}}
    \subfigure[The invariant mass of the two leading jet]{
    \label{distributionsemib}
    \includegraphics[width=0.46\textwidth]{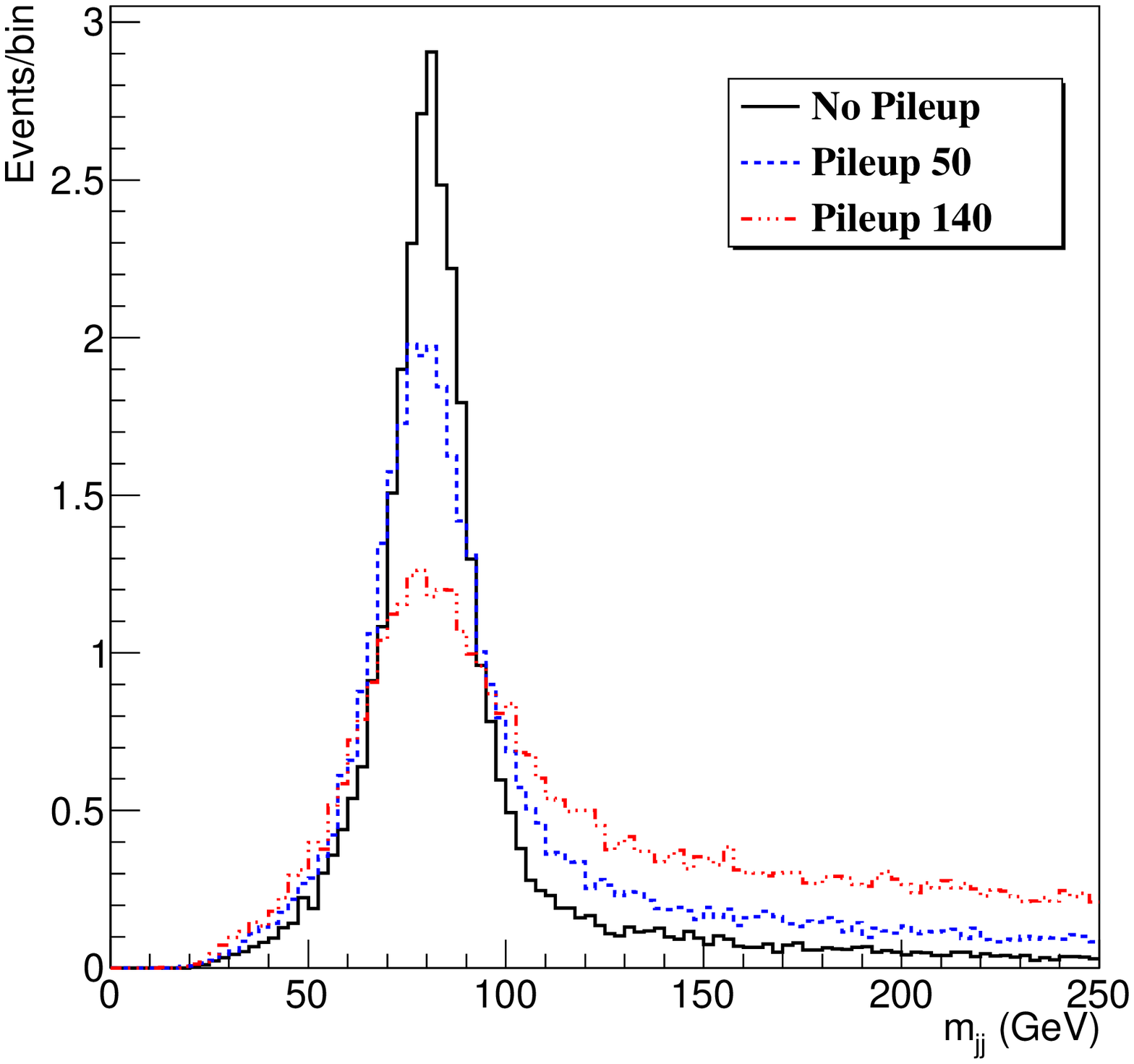}}

\caption{\label{diagrams} Distribution of signal process \ssl\ in different pileup scenarios at 14 TeV LHC}}
\end{figure}

Table.~\ref{tab5} shows the 8 TeV event numbers for the signal and backgrounds and significances and 14 TeV in Table.~\ref{tab6}. A significance of 0.5 $\sigma$ can be reached to observe the \www\ production in semileptonic decay channel at 8 TeV LHC and $0.91\sim 1.96 \sigma $ at 14 TeV. As shown in the Table.~\ref{tab6}, with the pileup increasing, the significance drops rapidly when using cut-based method. It mainly dues to the pileup jets which result in worse jet energy resolution. Fig.~\ref{distributionsemia} shows the jet number distribution from different pileup scenarios. Jet number increases with pileup. Especially, in 140 pileup scenario, most of events contain at least 4 jets, this would make the $N_{jet}$ cut has less efficiency to separate signal from backgrounds. Similarly, in Fig.~\ref{distributionsemib}, the invariant mass of two leading jets $m_{jj}$ distribution has a broader peak near W boson mass and harder tail in 140 pileup case. This would also reduce the discrimination between signal and backgrounds. In general, unlike the pure leptonic decay channel case, the significance of observing the $WWW$ production in semileptonic channel suffers more contamination from pileup event because it is difficult to identify the jet's original source, whether it come from signal or pileup.

		\begin{table}[h]
		  \centering
		    \begin{tabular}{c||c||c|c}
		    \hline
	        \multirow{2}{*}{Processes} & \multirow{2}{*}{Cross section[fb]} &\multicolumn{2}{c}{Events} \\\cline{3-4}
	   &  & cut-based & BDT \\
			\hline
		    $WWW$  & 1.732  & 1.9   & 1.7 \\
		    \hline
		    $t\bar{t}W$ & 169.7  & 1.7   & 1.6 \\\hline
		    $WWjj$  & 19.95  & 3.2   & 2.9 \\\hline
		    $WZjj$  & 192.3  & 8.4   & 4.9 \\\hline
			\hline
		     \multicolumn{2}{c||}{Significance} & 0.51  & 0.54  \\
		    \hline
		    \end{tabular}%
		  \caption{\label{tab5} Event numbers and significances of $WWW$ production in semileptonic decay channel at the LHC with $\sqrt{s}=8$ TeV in mean pileup 20 scenario with  integrated luminosity of $20\fbinv$.}
		\end{table}%

		\begin{table}[h]
		  \centering
		    \begin{tabular}{c||c||c|c|c|c|c|c}
		    \hline
		   \multirow{3}{*}{Processes} & \multirow{3}{*}{Cross section[fb]} & \multicolumn{6}{c}{Events}\\\cline{3-8}  
		 & &  \multicolumn{2}{c|}{Pileup 0} & \multicolumn{2}{c|}{Pileup 50} & \multicolumn{2}{c}{Pileup 140} \\\cline{3-8}
		    
		         & &  cut-based & BDT & cut-based & BDT & cut-based & BDT \\
		    \hline
		    $WWW$  & 3.586 & 22.1  & 21.9  & 22    & 20.7  & 21.2  & 39.7 \\
		    \hline
		    $t\bar{t}W$ & 480.2 & 14.4  & 19.7  & 53.2  & 36.5  & 112.6 & 140.2 \\\hline
		    $WWjj$ & 49.15 & 13.3  & 15.5  & 24.4  & 27.9  & 46.7  & 121.8 \\\hline
		    $WZjj$ & 627.9& 106.7 & 82.9  & 212.7 & 138.8 & 379.5 & 680 \\
		    \hline\hline
		    \multicolumn{2}{c||}{Significance}  & 1.86  & 1.96  & 1.28  & 1.43  & 0.91  & 1.28  \\
		    \hline
		    \end{tabular}%
		    	  \caption{\label{tab6} Event numbers and significances of $WWW$ production in semileptonic decay channel at the LHC with $\sqrt{s}=14$ TeV with integrated luminosity of $100\fbinv$.}
		  
		\end{table}%

\section{Anomalous $WWWW$ Couplings}
\label{anowww}
\subsection{aQGC in pure leptonic decay channel}
\label{14aqgc}
\begin{figure}{
\centering
\includegraphics[width=0.46\textwidth]{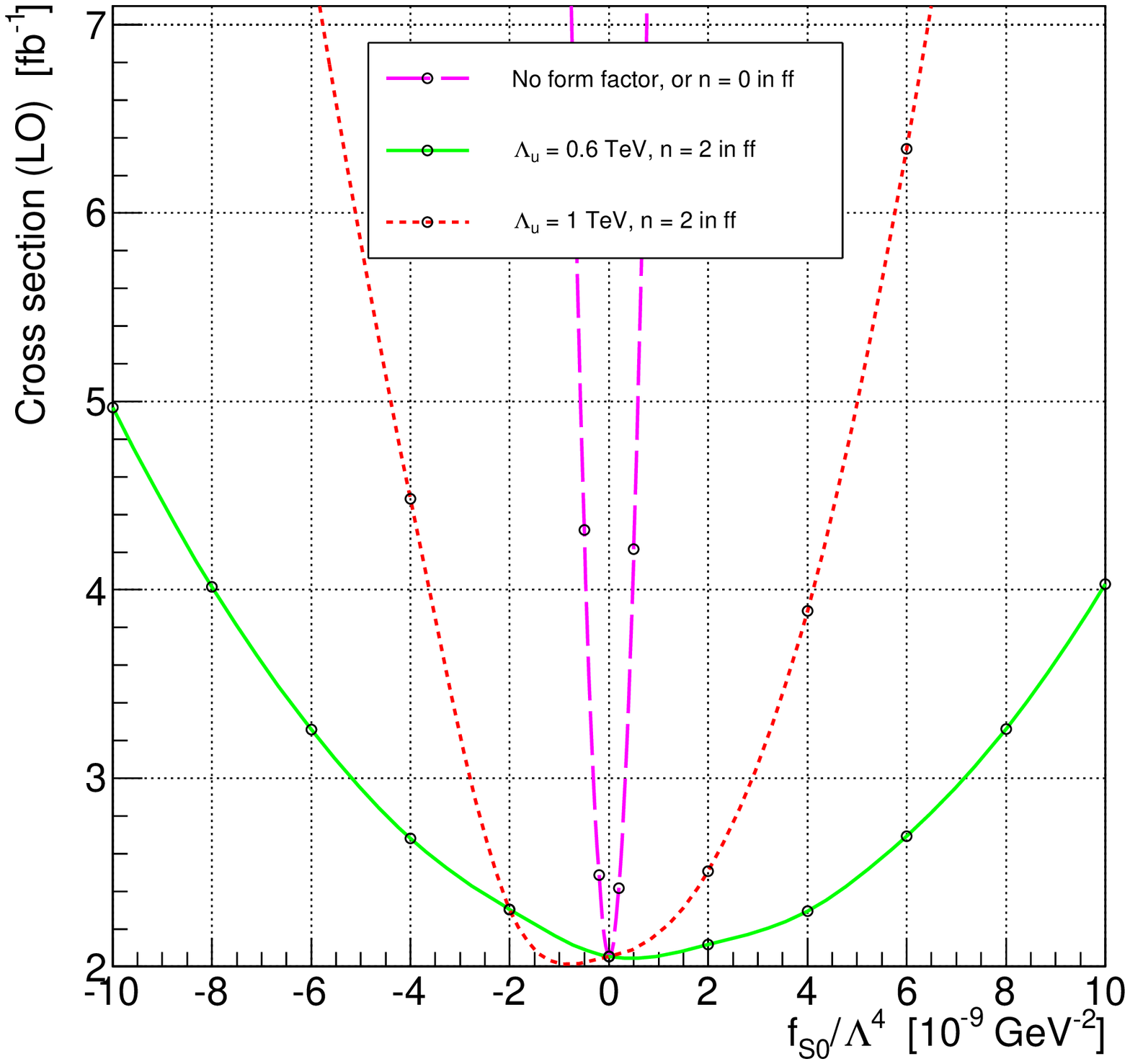}
\includegraphics[width=0.46\textwidth]{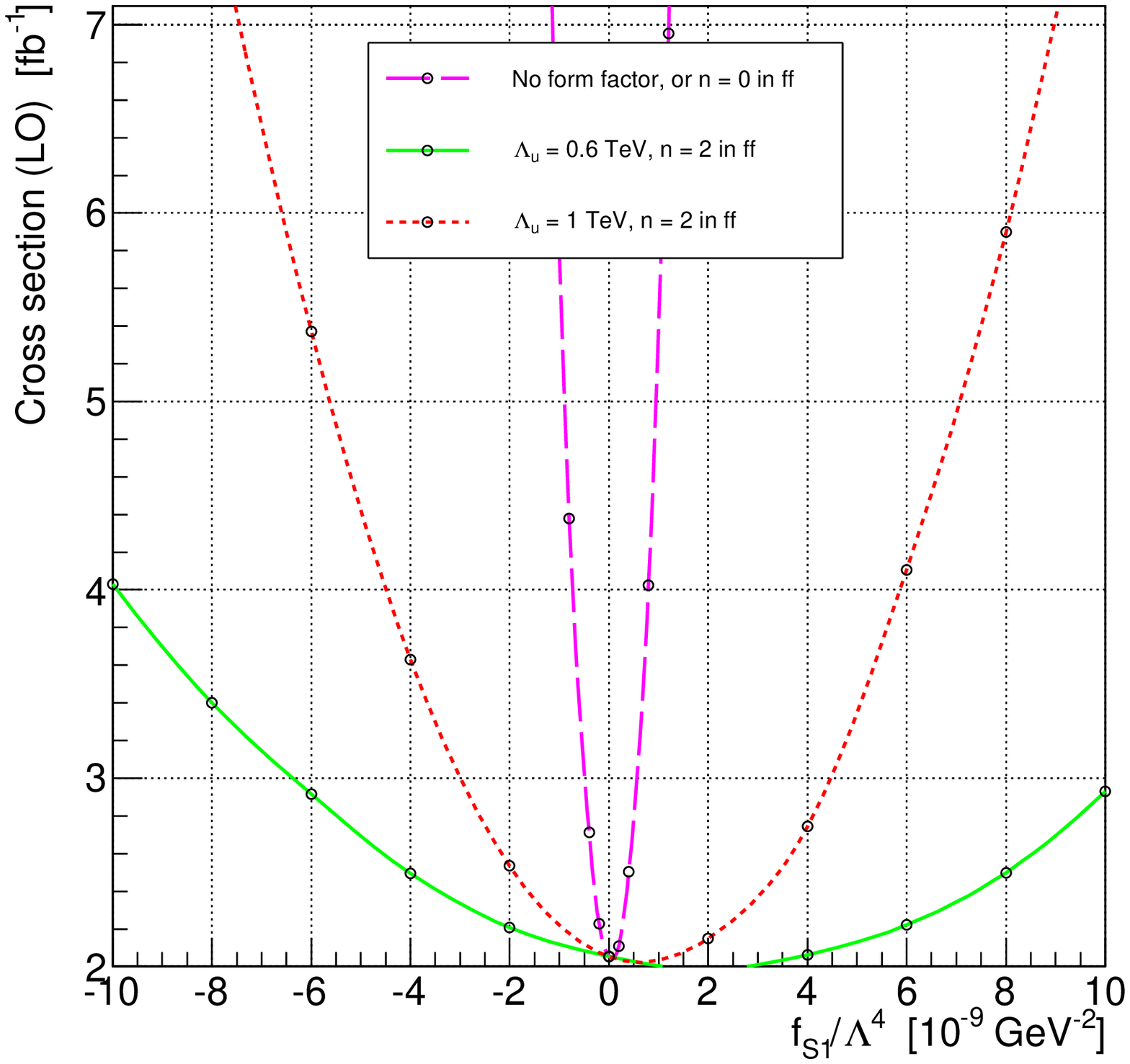}
\includegraphics[width=0.46\textwidth]{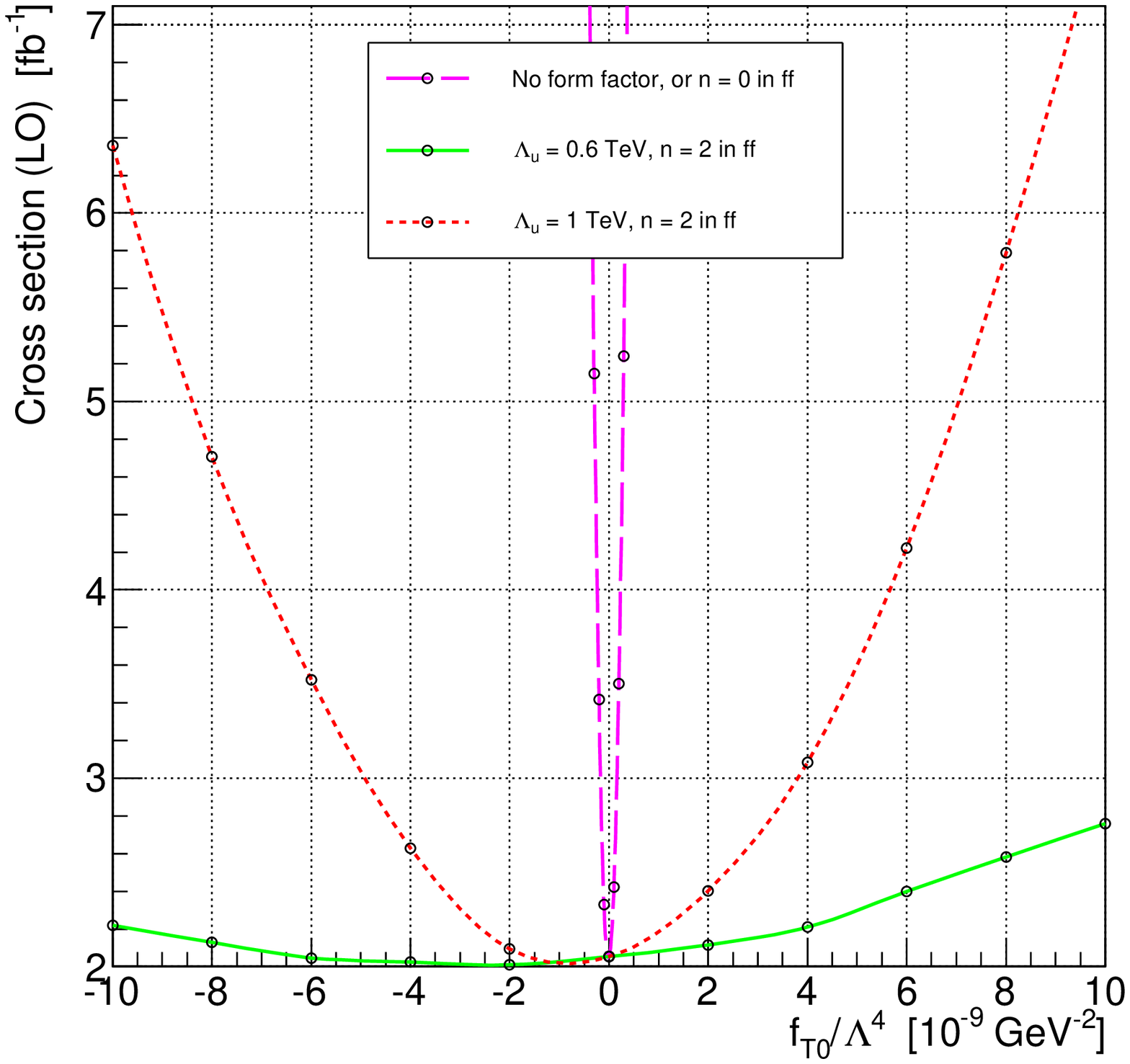}
\caption{\label{aQGCXS}Cross section dependences on $WWWW$ anomalous couplings $f_{S0,S1}/\Lambda^4$ and $f_{T0}/\Lambda^4$ at 14 TeV LHC with different form factors applied.}}
\end{figure}

\begin{figure}{
\centering
\includegraphics[width=0.49\textwidth]{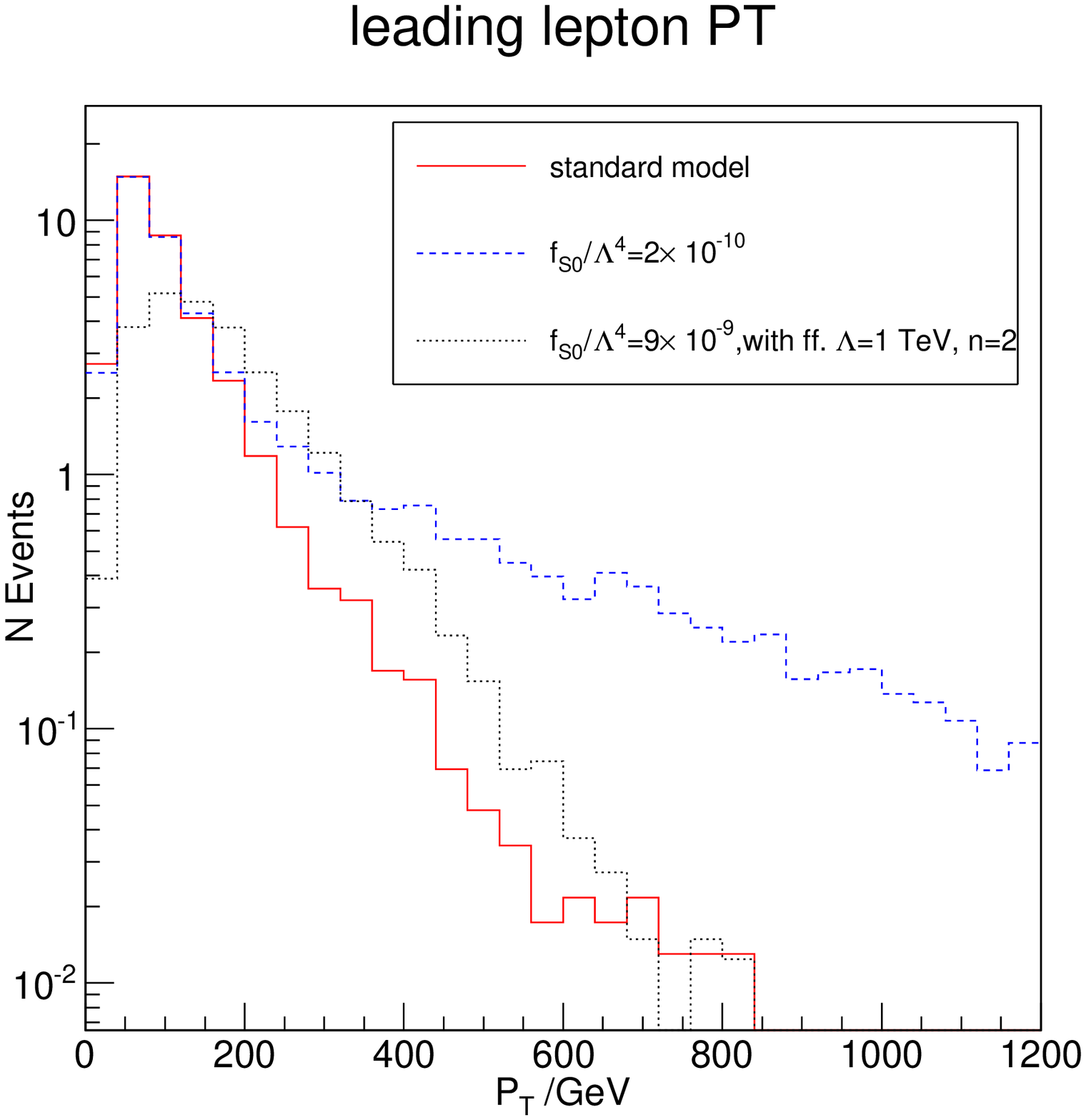}
\includegraphics[width=0.49\textwidth]{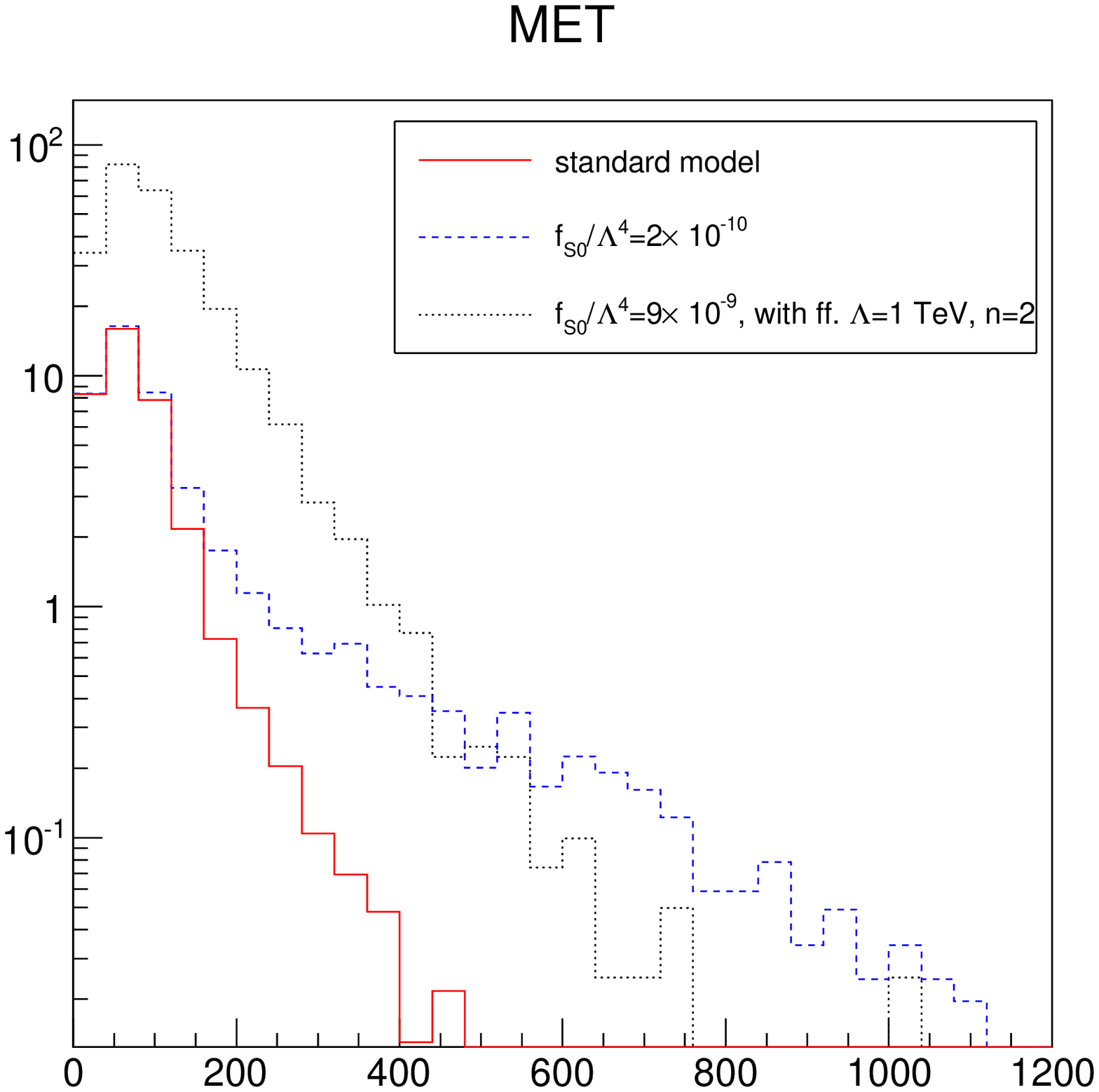}
\caption{\label{aQGC}The leading lepton $P_T$ and \met\ distributions for \www productions at 14 TeV LHC, with or without aQGCs and form factor.}}
\end{figure}

The $W^{\pm}W^{\pm}W^{\mp}$ production can be sensitive to aQGC $WWWW$. The cross sections, via \mgme\ after preselection cuts mentioned in Sec.~\ref{fulllep}, can grow quickly with the increase of the absolute values of aQGCs, as demonstrated in Fig.~\ref{aQGCXS}. Furthermore, as shown in Fig.~\ref{aQGC}, the aQGCs lead to excesses on the hard tails in various kinematic region. Thus we refine the cuts in both schemes in Sec.~\ref{fulllep} to enhance the sensitivity of QGCs without form factor as following, e.g.:
\begin{itemize}{
\item (1) $\met > 350$ GeV,
\item (2) The transverse mass of 3 leptons $m_T>1$ TeV,
\item (3) leading lepton $P_T>200$ GeV.
}
\end{itemize}

For the form factor case, \met\ and leptons would be softer, thus we refine our cuts as:

\begin{itemize}{
\item (1) $\met > 80$ GeV,
\item (2) The transverse mass of 3 leptons $m_T>250$ GeV,
\item (3) leading lepton $P_T>50$ GeV.

}
\end{itemize}


After all these selection cuts, the significances are calculated and displayed as following as functions of the QGCs $f_{S0}$, $f_{S1}$ and $f_{T0}$, at the 8 TeV LHC with an integrated luminosity of 20 fb$^{-1}$ and 14 TeV LHC with an integrated luminosity of 100 fb$^{-1}$, respectively. As shown in Ref.~\cite{snowmass:2013}, pileup would not affect aQGC measurement at hard kinematic region, thus we produce aQGC samples without pileup mixing. As mentioned in sec.~\ref{fulllep}, we category 2 different analysis schemes, but only the more stringent results (Scheme 2) are presented here. The results of 8 TeV LHC and 14 TeV LHC are shown in Table.~\ref{aQGC8TeV} and Table.~\ref{aQGC14TeV}, both with/without form factor results are shown. Compared with the Fig.~\ref{operators}, for 8 TeV results, even with form factors applied, the aQGC limits are unitarity unsafe. But for 14 TeV results, they are close to unitarity safe region.

  \begin{table}[h]
  \centering
  {
    \begin{tabular}{c|cc|cc}
    \hline
                & \multicolumn{2}{c|}{No form factor} & \multicolumn{2}{c}{$\Lambda=1$\,TeV, n=2}  \\\cline{2-5}

        & lower limit & upper limit& lower limit & upper limit\\\hline
     $\frac{f_{S0}}{\Lambda^4}$ & $-1.32\times 10^{-9}$  & $1.30\times 10^{-9}$   & $-8.74\times10^{-9}$ & $8.87\times10^{-9}$ \\
         $\frac{f_{S1}}{\Lambda^4}$  & $-2.00\times10^{-9}$  & $2.03\times10^{-9}$   & $-1.08\times 10^{-8}$ & $1.17\times10^{-8}$ \\
        $\frac{f_{T0}}{\Lambda^4}$ & $-5.56\times10^{-12}$  & $5.44\times10^{-12}$   &$-1.30\times10^{-10}$ & $1.21\times10^{-10}$ \\
          \hline
    
    \hline
    \end{tabular}%
    }
  \caption{Constraints on anomalous quartic couplings parameters $f_{S0}/\Lambda^4$, $f_{S1}/\Lambda^4$ and $f_{T0}/\Lambda^4$ at 8 TeV LHC via $WWW$ production pure leptonic decay channel with integrated luminosity of 20 fb$^{-1}$. Units are in GeV$^{-4}$.}
  \label{aQGC8TeV}%
\end{table}%

\begin{table}[h]
  \centering
  {
    \begin{tabular}{c|cc|cc|cc}
    \hline
                & \multicolumn{2}{c|}{No form factor} & \multicolumn{2}{c|}{$\Lambda=1$\,TeV, n=2} & \multicolumn{2}{c}{$\Lambda=0.5$\,TeV, n=2} \\\cline{2-7}

        & lower limit & upper limit& lower limit & upper limit & lower limit& upper limit\\\hline
     $\frac{f_{S0}}{\Lambda^4}$ & $-1.78\times 10^{-10}$  & $1.79\times 10^{-10}$   & $-2.80\times10^{-9}$ & $3.08\times10^{-9}$  & $-1.21\times10^{-8}$ & $1.29\times10^{-8}$ \\
         $\frac{f_{S1}}{\Lambda^4}$  & $-2.66\times10^{-10}$  & $2.78\times10^{-10}$   & $-3.47\times 10^{-9}$ & $4.44\times10^{-9}$  & $-1.29\times10^{-8}$ & $1.81\times10^{-8}$ \\
        $\frac{f_{T0}}{\Lambda^4}$ & $-5.80\times10^{-13}$  & $5.87\times10^{-13}$   & $-4.48\times10^{-11}$ & $3.46\times10^{-11}$   & $-2.46\times10^{-10}$  & $1.76\times10^{-10}$ \\
          \hline
    
    \hline
    \end{tabular}%
    }
  \caption{Constraints on anomalous quartic couplings parameters $f_{S0}/\Lambda^4$, $f_{S1}/\Lambda^4$ and $f_{T0}/\Lambda^4$ at 14 TeV LHC via $WWW$ production pure leptonic decay channel with integrated luminosity of 100 fb$^{-1}$. Units are in GeV$^{-4}$.}
  \label{aQGC14TeV}%
\end{table}%

\subsection{aQGC in semileptonic decay channel}
\label{14semiaqgc}
\begin{figure}{
\centering
\includegraphics[width=0.49\textwidth]{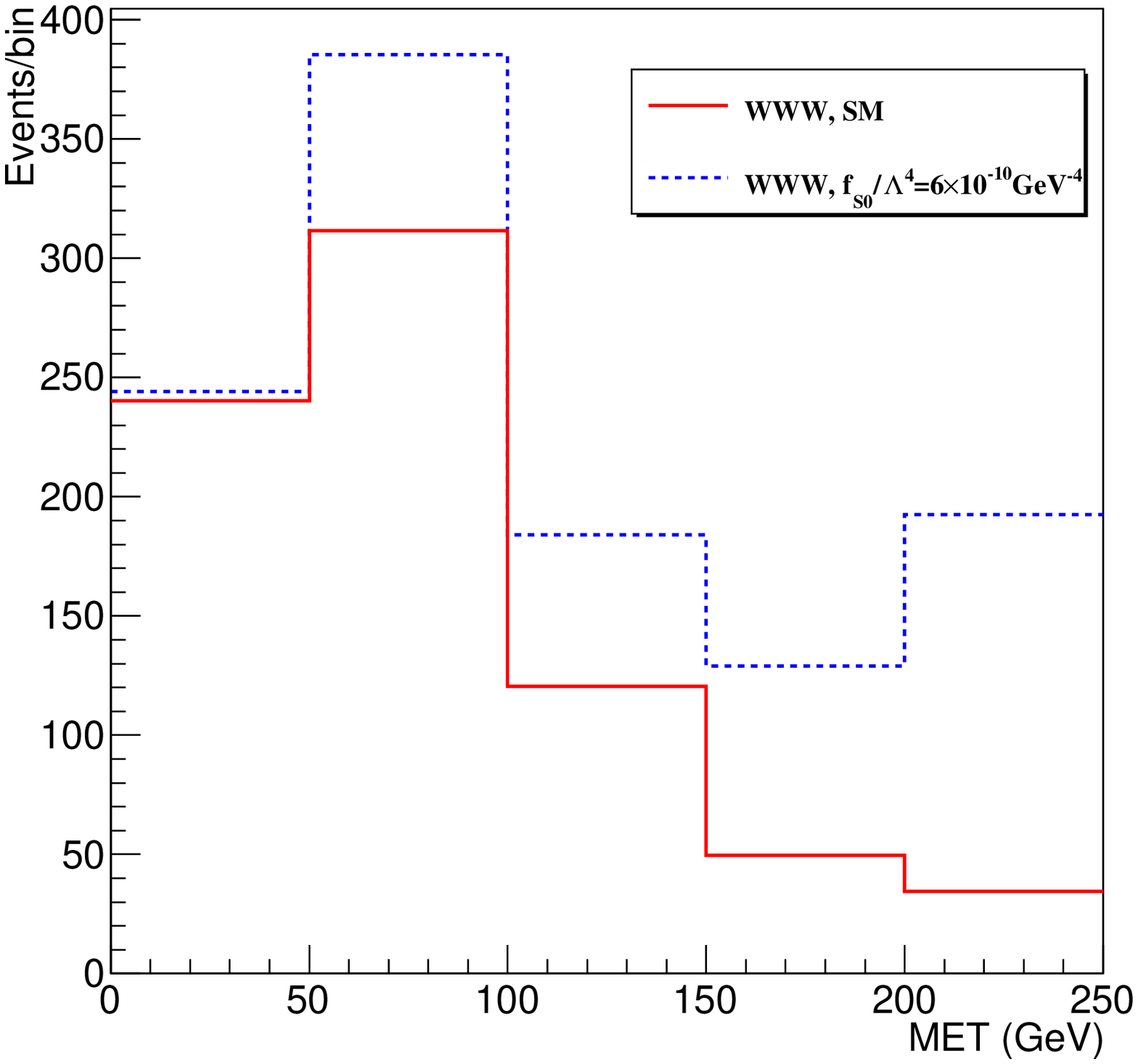}
\includegraphics[width=0.49\textwidth]{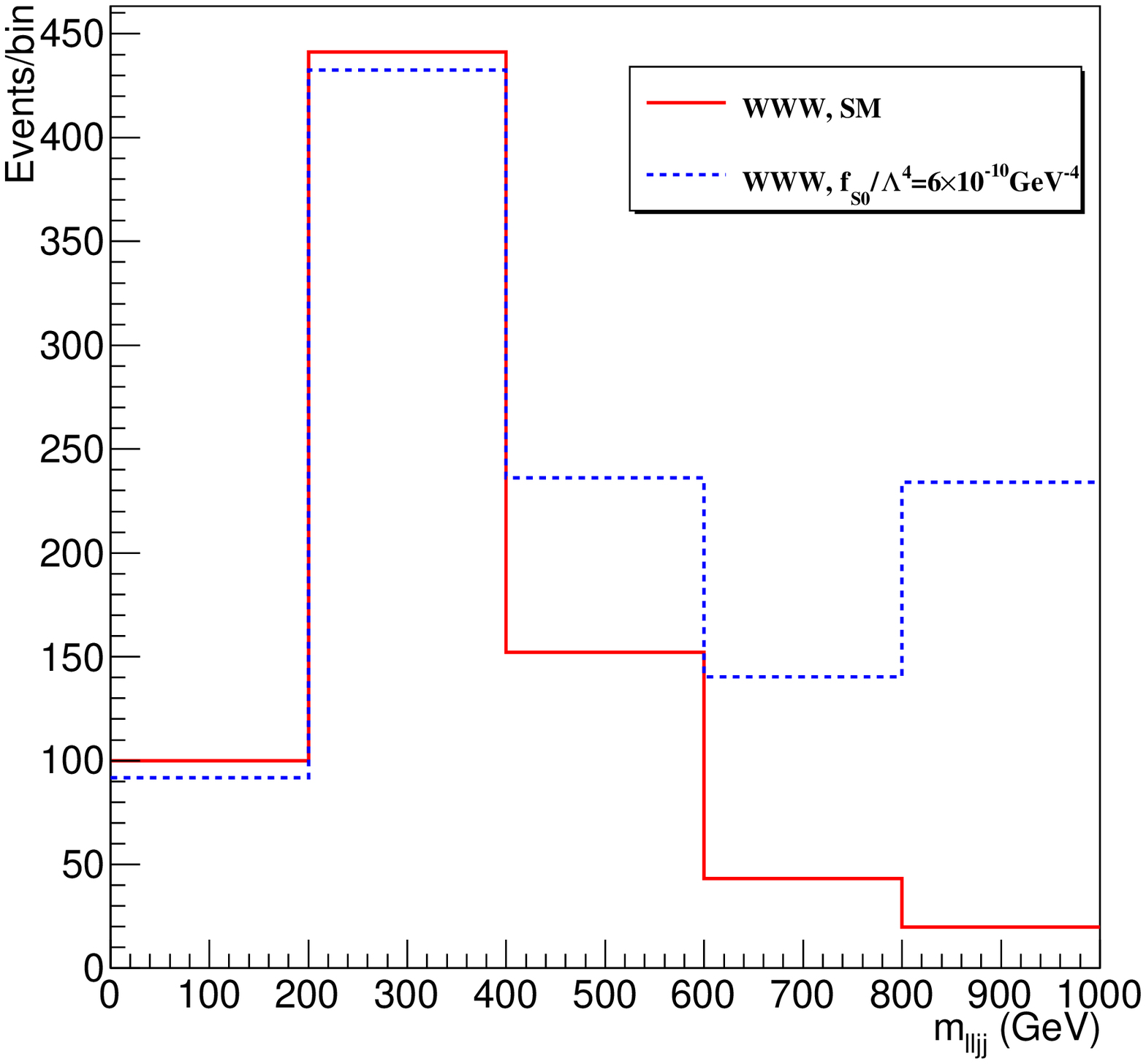}
\caption{\label{aQGCdistributionsemi}The in \met\ and $m_{lljj}$ distributions for \www productions at 14 TeV LHC, with the aQGCs $f_{S0}/\Lambda^4=6\times10^{-10}$GeV$^{-4}$. No form factor is applied here. The last bin includes overflow.}}
\end{figure}

After generating event and applying preselection cuts mentioned in Sec.~\ref{semilep} , some distributions of $WWW$ production and aQGC of $f_{S0}/\Lambda^4=6\times10^{-10}$ GeV at 14 TeV LHC are shown in Fig.~\ref{aQGCdistributionsemi}. The aQGC has more excess at hard tail. Based on this characteristic, to further improve the sensitivity on aQGC, we refine the cuts in addition to those cuts in Sec.~\ref{semilep}:

\begin{itemize}{
\item (1) The invariant mass of the same sign lepton pair plus two leading jets $m_{lljj}\geq 600$ GeV,
\item (2) $\met > 150$ GeV.
}
\end{itemize}

For the form factor case, \met\ and jets would be softer, thus we refine our cuts as:

\begin{itemize}{
\item (1) The invariant mass of the same sign lepton pair plus two leading jets $m_{lljj}\geq 200$ GeV,
\item (2) $\met > 50$ GeV.
}
\end{itemize}

The aQGC limits for 14 TeV LHC are given in Table.~\ref{aQGCsemi14TeV}, via $WWW$ production semileptonic decay channel with integrated luminosity of 100 fb$^{-1}$, with/without form factor.

One can also compare our results with the previous MC simulation given by Snowmass Collaboration~\cite{snowmass:2013} and O. Eboli et.al. based on vector boson fusion(VBF)~\cite{VBFWWWW:2006}, respectively, as shown in Table ~\ref{tabcom}. The semileptonic channel results still suffer from bad jet energy resolution. But in pure leptonic channel, due to the optimized selection, we set a more stringent limit on $f_{T0}/\Lambda^4$, $8\times10^{-13}$GeV$^{-4}$ in 5$\sigma$ with 100\fbinv. This result is better than Snowmass one. As to the weak boson fusion, however, our \www\ channel seems to set looser limits on QGCs. However, triple W production channel has simpler event topology and populates at different kinematic phase space, thus can present us more information other than VBF channel.

\begin{table}[h]
  \centering
  {
    \begin{tabular}{c|cc|cc|cc}
    \hline
                & \multicolumn{2}{c|}{No form factor} & \multicolumn{2}{c|}{$\Lambda=1$\,TeV, n=2} & \multicolumn{2}{c}{$\Lambda=0.5$\,TeV, n=2} \\\cline{2-7}

        & lower limit & upper limit& lower limit & upper limit & lower limit& upper limit\\\hline
     $\frac{f_{S0}}{\Lambda^4}$ & $-4.56\times 10^{-10}$  & $4.58\times 10^{-10}$   & $-3.08\times10^{-9}$ & $3.39\times10^{-9}$  & $-1.20\times10^{-8}$ & $1.40\times10^{-8}$ \\
         $\frac{f_{S1}}{\Lambda^4}$  & $-9.46\times10^{-10}$  & $9.85\times10^{-10}$   & $-4.00\times 10^{-9}$ & $5.26\times10^{-9}$  & $-1.28\times10^{-8}$ & $1.77\times10^{-8}$ \\
        $\frac{f_{T0}}{\Lambda^4}$ & $-2.80\times10^{-12}$  & $2.70\times10^{-12}$   & $-7.60\times10^{-11}$   & $6.00\times10^{-11}$    & $-4.03\times 10^{-10}$  & $2.88\times10^{-10}$ \\
          \hline

    \end{tabular}%
    }
  \caption{Constraints on anomalous quartic couplings parameters $f_{S0}/\Lambda^4$, $f_{S1}/\Lambda^4$ and $f_{T0}/\Lambda^4$ at 14 TeV LHC via $WWW$ production semileptonic decay channel with integrated luminosity of 100 fb$^{-1}$. Units are in GeV$^{-4}$.}
  \label{aQGCsemi14TeV}%
  
\end{table}%

\begin{table}[!ht]
{
\begin{footnotesize}
\centering{
\begin{tabular}{c|c|c|c|c|c|c|c|c}\hline
          & \multicolumn{2}{c|}{$WWW$ in p.l decay} & \multicolumn{2}{c|}{$WWW$ in s.l decay}  &\multicolumn{2}{c|}{ VBF $WW$ } & \multicolumn{2}{c}{Snowmass $WWW$}  \\
          &  \multicolumn{2}{c|}{95\% CL with 100\fbinv}    & \multicolumn{2}{c|}{95\% CL with 100\fbinv}&   \multicolumn{2}{c|}{99\% CL with 100\fbinv}  &  \multicolumn{2}{c}{5$\sigma$ with 300\fbinv}  \\\hline   
           & lower limit & upper limit& lower limit & upper limit & lower limit& upper limit& lower limit& upper limit\\\hline                    
 $\frac{f_{S0}}{\Lambda^4}$ &$-1.8\times 10^{-10}$ &$1.8\times 10^{-10}$ &$-4.6\times10^{-10}$ &$4.6\times10^{-10}$&$-2.2\times 10^{-11}$ &$2.4\times 10^{-11}$ & -     &- \\\hline
 $\frac{f_{S1}}{\Lambda^4}$  &$-2.7\times 10^{-10}$ &$2.8\times 10^{-10}$ &$-9.5\times10^{-10}$&$9.9\times10^{-10}$ &$-2.5\times 10^{-11}$  &$2.5\times 10^{-11}$ & -                        &- \\\hline
 $\frac{f_{T0}}{\Lambda^4}$  &$-5.8\times 10^{-13}$ &$5.9\times 10^{-13}$ &$-2.8\times10^{-12}$  &$2.7\times10^{-12}$ &- &  - &- &$1.2\times 10^{-12}$      \\\hline
\end{tabular}}
\end{footnotesize}
\caption{\label{tabcom} Constraints on aQGC parameter limit comparison to previous MC study at 14 TeV LHC. All without form factor applied. Units are in GeV$^{-4}$.}}
\end{table}

\section{$WWW$ production and aQGC at 100 TeV future pp collider}
\label{100TeV}
We also studied $WWW$ production and aQGCs at 100 TeV future proton-proton collider. The simulation framework is basically the same as in 8 TeV and 14 TeV LHC study. However, at the level of detetor fast simulation , we use the Snowmass combined LHC detector which is a hybrid of CMS and ATLAS detectors~\cite{Snowmass:detector}, using the tracker components from CMS and calorimeter from ATLAS, etc. Effects of average 50 and 140 pileup scenarios will be considered here. 

\subsection{Pure leptonic decay channel}

For SM \www production and aQGC in pure leptonic decay, event selection cuts are the same as the studies of LHC in Sec.~\ref{fulllep} and~\ref{anowww}. The event numbers for the signal, backgrounds and significances are listed in Table.~\ref{100TeVsmpure}. One can see that it reaches a significance of $10\sim14\sigma$ to observe SM \www production at 100 TeV future proton-proton collider with 3000\fbinv integrated luminosity. 

\begin{table}[h!]
\centering
\begin{tabular}{c||c||c|c|c|c}
\hline
\multirow{4}{*}{Processes} & \multirow{4}{*}{Cross section[fb]} & \multicolumn{4}{c}{ Events } \\
\cline{3-6}
& & \multicolumn{4}{c}{cut-based}  \\
\cline{3-6}
& &  \multicolumn{2}{c|}{Pileup 50} &\multicolumn{2}{c}{Pileup 140} \\
\cline{3-6}
& & s1& s2 &s1& s2 \\
\hline
$WWW$        & 15.6 & 4758   &  1416   &   3855  &   1156     \\\hline
$WZ$         & 2570 &  92185  &  1670    &  82060   &   1696    \\\hline
$t\bar{t}W$  & 89.7 & 8607   &  2539   &   9930   &     3211    \\\hline
$ZZ$         & 2674 &  26633  &  481  &  24226  &      1283      \\\hline
$t\bar{t}Z$  & 454&  15240  &  4408  &   18180   &  5034          \\\hline
$WWZ$        & 14.1 &  1164  &  317   &  993    &   255    \\\hline\hline
 \multicolumn{2}{c||}{Significance}& 12.5   & 14.6 & 10.5 & 10.8 \\\hline

\end{tabular}
\caption{\label{100TeVsmpure} Event numbers and significances of $WWW$ production in pure leptonic decay channel at future proton-proton collider with $\sqrt{s}=100$ TeV and integrated luminosity of $3000$ fb$^{-1}$.}
\end{table}

\subsection{Semileptonic decay channel}
Unlike the LHC, in 100 TeV proton proton collider, the pileup contamination in semileptonic channel is more severe. Therefore, we optimise the event selection cuts again(most of them are related to jets)(See table.~\ref{cut100TeV}). We only consider those jet in the tracker region, namely, $|\eta|\leq2.5$. For this, we define $N^{tight}_{jet}$ as the number of jets which satisfy $p_T\geq30$ GeV and $|\eta|\leq2.5$.  The results are shown in Table.~\ref{100TeVsmsemi}. It reaches $10\sim16\sigma$ to observe the SM \www production.

\begin{table}[h!]
  \centering
    \begin{tabular}{lcc}
    \hline
    Pileup   & 50    & 140 \\
	\hline
    \nlep  & \multicolumn{2}{c}{$=2$} \\
    
    Sign & \multicolumn{2}{c}{$(+,+)$ or $(-, -)$} \\
    \met   & \multicolumn{2}{c}{$\ge30$\,GeV} \\
    \njetTight  &  $\ge 2, \le 4$ & $\ge 2$ \\
    $|m_{jj}-m_W|$ & $\le25$\,GeV & $\le40$\,GeV \\
    \hline
    \end{tabular}%
  
  \caption{\label{cut100TeV}\ssl\ Event selections at 100 TeV proton proton collider}
\end{table}%

\begin{table}[h!]
  \centering
  {
    \begin{tabular}{c||c||c|c|c|c}
    \hline
   \multirow{3}{*}{Processes} & \multirow{3}{*}{Cross section[fb]} & \multicolumn{4}{c}{Events}\\\cline{3-6}  
		 &  & \multicolumn{2}{c|}{Pileup 50} & \multicolumn{2}{c}{Pileup 140} \\\cline{3-6}	    
		      
          &    & cut-based & BDT   & cut-based & BDT \\
    \hline
    $WWW$  &26 &  6465  & 12156 & 7794  & 13485 \\\hline
    $t\bar{t}W$ & 7684   & 35961 & 65928 & 60396 & 100047 \\\hline
    $WWjj$ & 535  & 30507 & 41124 & 71610 & 75708 \\\hline
    $WZjj$ & 16250 & 209820 & 437775 & 429195 & 693225 \\\hline
	\hline
     \multicolumn{2}{c||}{Significance}    & 12.3  & 16.4  & 10.4  & 14.4  \\
    \hline
    \end{tabular}%
    }
  \caption{\label{100TeVsmsemi} Event numbers and significances of $WWW$ production in semileptonic decay channel at future proton-proton collider with $\sqrt{s}=100$ TeV and integrated luminosity of $3000$ fb$^{-1}$.}
\end{table}%

\subsection{Anomalous quartic couplings}
\label{aQGC100TeV}
We also wish to explore the potential of probing aQGC at 100 TeV collider. The event selections of aQGC is basically the same as in Sec.~\ref{anowww}. We list the results of both pure leptonic and semileptonic channel in Table.~\ref{aQGCfull100TeV} and  Table.~\ref{aQGCsemi100TeV}

\begin{table}[h!]
  \centering{
\begin{tabular}{c|cc|cc|cc}
\hline
    & \multicolumn{2}{c|}{No form factor} & \multicolumn{2}{c|}{$\Lambda=1$\,TeV, n=2} & \multicolumn{2}{c}{$\Lambda=0.5$\,TeV, n=2} \\\cline{2-7}
    & lower limit & upper limit& lower limit & upper limit & lower limit& upper limit\\\hline
$\frac{f_{S0}}{\Lambda^4}$ &$-2.93\times10^{-12}$  & $3.04\times10^{-12}$ & $-1.65\times10^{-9}$ & $1.50\times10^{-9}$  & $-2.06\times10^{-8}$ & $2.15\times10^{-8}$ \\
$\frac{f_{S1}}{\Lambda^4}$ &$-1.30\times10^{-12}$  & $1.16\times10^{-12}$ & $-1.87\times10^{-9}$ & $2.37\times10^{-9}$  & $-2.75\times10^{-8}$ & $2.84\times10^{-8}$ \\
$\frac{f_{T0}}{\Lambda^4}$ &$-3.69\times10^{-15}$  & $2.97\times10^{-15}$ & $-9.18\times10^{-12}$& $6.76\times10^{-12}$ & $-9.90\times10^{-11}$  &$7.30\times10^{-11}$ \\\hline
    \end{tabular}%
    }
  \caption{Constraints on anomalous quartic couplings parameters $f_{S0}/\Lambda^4$, $f_{S1}/\Lambda^4$ and $f_{T0}/\Lambda^4$ at 100 TeV future proton proton collider via $WWW$ production pure leptonic decay channel with integrated luminosity of 3000 fb$^{-1}$. Units are in GeV$^{-4}$.}
  \label{aQGCfull100TeV}%
  
\end{table}%

\begin{table}[h!]
  \centering
  {
    \begin{tabular}{c|cc|cc|cc}
    \hline
                & \multicolumn{2}{c|}{No form factor} & \multicolumn{2}{c|}{$\Lambda=1$\,TeV, n=2} & \multicolumn{2}{c}{$\Lambda=0.5$\,TeV, n=2} \\\cline{2-7}

        & lower limit & upper limit& lower limit & upper limit & lower limit& upper limit\\\hline
     $\frac{f_{S0}}{\Lambda^4}$ & $-1.03\times 10^{-10}$  & $1.00\times 10^{-10}$   & $-8.79\times10^{-10}$ & $1.17\times10^{-9}$  & $-2.99\times10^{-9}$ & $5.18\times10^{-9}$ \\
         $\frac{f_{S1}}{\Lambda^4}$  & $-1.93\times10^{-10}$  & $2.21\times10^{-10}$   & $-1.08\times 10^{-9}$ & $2.27\times10^{-9}$  & $-3.26\times10^{-9}$ & $7.59\times10^{-9}$ \\
        $\frac{f_{T0}}{\Lambda^4}$ & $-2.00\times10^{-13}$  & $2.00\times10^{-13}$   & $-3.10\times10^{-11}$   & $1.60\times10^{-11}$    & $-1.84\times 10^{-10}$  & $6.80\times10^{-11}$ \\
          \hline

    \end{tabular}%
    }
  \caption{Constraints on anomalous quartic couplings parameters $f_{S0}/\Lambda^4$, $f_{S1}/\Lambda^4$ and $f_{T0}/\Lambda^4$ at 100 TeV proton proton collider via $WWW$ production semileptonic decay channel with integrated luminosity of 3000 fb$^{-1}$. Units are in GeV$^{-4}$.}
  \label{aQGCsemi100TeV}%

\end{table}%

\section{Unitarity safety discussion}
\label{UV}
In the previous study, besides no from factor case, we exploit also form factors with $\Lambda_{ff}=0.5 / 1$ TeV and $n=2$ TeV. As for both 14 and 100 TeV future collider, by comparing Tables.~\ref{aQGC14TeV}, \ref{aQGCsemi14TeV}, \ref{aQGCfull100TeV}, \ref{aQGCsemi100TeV} with Fig.~\ref{operators}, one can see that applying those form factors will not yet lead to unitarity safe, due to lack of luminosity. We provide the needed luminosity (estimated based on our analysis in Sec.~\ref{aQGC100TeV} to reach unitarity safety in Tables.~\ref{safeuni1TeV} and \ref{safeunip5TeV}. In general, it needs very high luminosity to reach safe unitarity.  Same situation appears in the 8TeV CMS $WV\gamma$ measurement~\cite{WVgammaCMS:2014} which finds that the unitarity safety can not be satisfied with a dipole form factor, however, "unitarity conserving new physics with a structure more complex than that represented by a dipole form factor is possible"~\cite{WVgammaCMS:2014} .

\begin{table}[h!]
  \centering
  {
    \begin{tabular}{c|c}
    \hline
    aQGC, $\Lambda=1$\,TeV, n=2 & Required Luminosity($fb^{-1}$) \\\hline
    $\frac{f_{S0}}{\Lambda^4}=4\times 10^{-11} $ & 16000\\\hline
    $\frac{f_{S1}}{\Lambda^4}=1\times 10^{-11} $ & 19000\\\hline
    $\frac{f_{T0}}{\Lambda^4}=1\times 10^{-12} $ & 4500 \\\hline
    \end{tabular}%
    }
  \caption{The unitarity safe aQCG boundary ( from Fig.~\ref{operators}) and corresponding required luminosity to reach it, for 14TeV future collider, with form factor $\Lambda=1$ TeV and $n=2$ }
  \label{safeuni1TeV}%

\end{table}%

\begin{table}[h!]
  \centering
  {
    \begin{tabular}{c|c}
    \hline
    aQGC, $\Lambda=0.5$\,TeV, n=2 & Required Luminosity($fb^{-1}$) \\\hline
     $\frac{f_{S0}}{\Lambda^4}=6\times 10^{-10} $ & 6000\\\hline
     $\frac{f_{S1}}{\Lambda^4}=3\times 10^{-10} $ & 12000\\\hline
     $\frac{f_{T0}}{\Lambda^4}=2\times 10^{-11} $ & 4000 \\\hline
    \end{tabular}%
    }
  \caption{The unitarity safe aQCG boundary ( from Fig.~\ref{operators}) and corresponding required luminosity to reach it, for 14 TeV future collider, with form factor $\Lambda=0.5$ TeV and $n=2$  }
  \label{safeunip5TeV}%

\end{table}%

\section{Conclusion}
\label{conclu}

The future upgrade of LHC and the next generation 100 TeV proton-proton collider with higher center of mass energy and luminosity enable measurement of triple gauge boson production and anomalous quartic gauge couplings, and \www\ production will be a potential process that can be exploited to test the SM predictions and probe $WWWW$ anomalous coupling exclusively with lower background contamination.

In summary, our study shows that at 8 TeV LHC with an integrated luminosity of 20 fb$^{-1}$ , 14 TeV LHC with 100 fb$^{-1}$ and 100 TeV next generation proton-proton collider with 3000\fbinv, for pure(semi-) leptonic decay channel one can reach a significance of about 0.4(0.5), 1.2(2) and 10(14) $\sigma$ to probe the SM \www\ production, and can constrain at 95\% C.L. the anomalous $WWWW$ coupling parameters $f_{S0,S1}/\Lambda^4$ at $1\times 10^{-9}$ GeV$^{-4}$ and $f_{T0}/\Lambda^4$ at $1\times 10^{-12}$ GeV$^{-4}$ at 8 TeV LHC with 20 fb$^{-1}$, $f_{S0,S1}/\Lambda^4$ at $1\times 10^{-10}(10^{-10})$ GeV$^{-4}$ and $f_{T0}/\Lambda^4$ at $1\times 10^{-13}(10^{-12})$ GeV$^{-4}$ at 14 TeV LHC with 100 fb$^{-1}$ and $f_{S0,S1}/\Lambda^4$ at $1\times 10^{-12}(10^{-10})$ GeV$^{-4}$ and $f_{T0}/\Lambda^4$ at $1\times 10^{-15}(10^{-13})$ GeV$^{-4}$ at 100 TeV with 3000 fb$^{-1}$. When the energy scale moving from 14 TeV to 100 TeV, the significance gain in pure leptonic channel is bigger than semileptonic channel, which may be due to that the QCD backgrounds increase much faster than the pure leptonic background. 

Our limits on $\cal L_{S}$ operators are presented for the first time in \www\ channel, although less tighter than the previous results from VBF process, however, triple W production channel populates at different kinematic phase space, thus can present us more information other than VBF channel. On the other hand, our limits on $f_{T0}/\Lambda^4$ are better than Snowmass due to optimized selection cuts. Moreover, it is the first time to study the \www\ production in semileptonic channel.

\acknowledgments
This work is supported in part by the National Natural Science Foundation of China, under Grants No. 11475180,
, No. 10721063, No. 10975004, No. 10635030 and No. 11205008, and National Fund for Fostering Talents in Basic Science, under Grant No. J1103206.

\end{document}